\documentclass[screen]{acmart}
\AtBeginDocument{%
  }

\setcopyright{acmlicensed}
\copyrightyear{2018}
\acmYear{2018}
\acmDOI{XXXXXXX.XXXXXXX}
\acmConference[Conference acronym 'XX]{Make sure to enter the correct conference title from your rights confirmation email}{June 03--05, 2018}{Woodstock, NY}
\acmISBN{978-1-4503-XXXX-X/2018/06}

\usepackage{amsmath}
\usepackage{graphicx}
\usepackage[ruled,vlined,linesnumbered]{algorithm2e}
\usepackage{subcaption}
\usepackage{xcolor}
\usepackage{hyperref}
\usepackage{xspace}
\usepackage{wasysym}
\usepackage{pifont}
\usepackage{multirow}
\usepackage{multicol}
\usepackage{booktabs}
\usepackage[table]{xcolor}
\usepackage{colortbl}
\usepackage{diagbox}
\usepackage{enumitem}
\usepackage{makecell}
\usepackage{array} 
\newcommand{\cmark}{\ding{51}}%
\newcommand{\xmark}{\ding{55}}%

\renewcommand{\sectionautorefname}{Section}

\def\eg{\textit{e.g.}\xspace}

\begin{document}

\title{A Comprehensive Survey of Website Fingerprinting Attacks and Defenses in Tor: Advances and Open Challenges}

\author{Yuwen Cui}
\email{ycui@usf.edu}
\orcid{0009-0003-3872-5445}
\affiliation{%
  \institution{University of South Florida}
  \city{Tampa}
  \state{FL}
  \country{USA}
}

\author{Guangjing Wang}
\affiliation{%
  \institution{University of South Florida}
  \city{Tampa}
  \state{FL}
  \country{USA}}
\email{guangjingwang@usf.edu}

\author{Khanh Vu}
\affiliation{%
  \institution{University of South Florida}
  \city{Tampa}
  \state{FL}
  \country{USA}}
\email{khanhduyvu@usf.edu}

\author{Kai Wei}
\affiliation{%
  \institution{University of South Florida}
  \city{Tampa}
  \state{FL}
  \country{USA}
}
\email{kwei@usf.edu}

\author{Kehan Shen}
\affiliation{%
  \institution{University of South Florida}
  \city{Tampa}
  \state{FL}
  \country{USA}
}
\email{kshen@usf.edu}

\author{Zhengyuan Jiang}
\affiliation{%
  \institution{University of South Florida}
  \city{Tampa}
  \state{FL}
  \country{USA}}
\email{jz1@usf.edu}

\author{Xiao Han}
\affiliation{%
  \institution{University of South Florida}
  \city{Tampa}
  \state{FL}
  \country{USA}}
\email{xiaoh@usf.edu}

\author{Ning Wang}
\affiliation{%
  \institution{University of South Florida}
  \city{Tampa}
  \state{FL}
  \country{USA}}
\email{ningw@usf.edu}

\author{Zhuo Lu}
\affiliation{%
  \institution{University of South Florida}
  \city{Tampa}
  \state{FL}
  \country{USA}}
\email{zhuolu@usf.edu}

\author{Yao Liu}
\affiliation{%
  \institution{University of South Florida}
  \city{Tampa}
  \state{FL}
  \country{USA}}
\email{yliu21@usf.edu}

\renewcommand{\shortauthors}{Yuwen Cui et al.}


\begin{abstract}

The Tor network provides strong anonymity by routing user traffic through multiple relays. However, it remains vulnerable to website fingerprinting (WF) attacks, which can accurately infer visited websites from encrypted traffic. To mitigate this threat, numerous defense mechanisms, including adaptive padding, traffic shaping, traffic morphing, and adversarial perturbation, have been proposed, although they often incur trade-offs between privacy and performance. Despite extensive research, a comprehensive survey that unifies the challenges and future directions of WF datasets, attack models, and defense mechanisms remains lacking. This paper addresses this gap by systematically reviewing these three key dimensions. We analyze representative approaches, compare their design principles, strengths, and limitations, and discuss emerging challenges, including multi-tab browsing, model drift, coarse-grained traffic representations, and realistic deployment. We also identify promising future research directions and provide a publicly available GitHub repository summarizing representative WF attacks, defenses, and datasets to facilitate reproducible research: \url{https://github.com/WF-Attack-and-Defense/Awesome-WF-Security}.
\end{abstract}






\maketitle

\section{Introduction}
\label{sec:intro}
The Tor network is a decentralized, volunteer-operated anonymity system that employs onion routing and layered encryption to conceal users’ identities and browsing activities. By routing traffic through multiple encrypted relay nodes, Tor is designed to resist surveillance, traffic analysis, and censorship, thereby preventing adversaries from linking users to their online activities or the websites they visit~\cite {dingledine2004tor, mani2018understanding, dantas2020detection, panchenko2011website}. Despite these protections, Tor remains vulnerable to website fingerprinting (WF) attacks~\cite{cherubin2022online, juarez2014critical, cai2012touching}, which analyze statistical characteristics of encrypted traffic to infer the websites users visit. Previous studies have shown that features such as packet burst patterns~\cite{herrmann2009website, wang2014effective, abe2016fingerprinting}, packet timing~\cite{panchenko2016website, rahman2019tik, yin2024traces}, and traffic directions~\cite{rahman2019tik, yan2018feature} can reveal distinctive website-specific patterns even when traffic content is encrypted. To mitigate this threat, researchers have proposed a wide range of defense mechanisms~\cite{cai2014systematic, juarez2016toward, nithyanand2014glove}, including traffic obfuscation~\cite{liu2023smart, shen2024real}, traffic regularization~\cite{holland2020regulator, cai2014systematic}, adversarial perturbation~\cite{hou2021universal, sun2022practical}, and website morphing~\cite{chan2018website, al2019bimorphing, ling2024wfguard}. While these defenses can substantially reduce fingerprinting effectiveness, they often incur additional bandwidth and latency overhead, creating a fundamental trade-off between anonymity protection and system performance.

\subsection{Motivation and Related Surveys}

Despite the growing significance of WF attacks and defenses, the field still lacks a systematic and comprehensive survey that synthesizes its rapidly evolving landscape. Existing surveys primarily focus on summarizing representative WF attack and defense mechanisms. For example, prior studies~\cite{karunanayake2021anonymisation, murad2026website, aminuddin2023rise} provide valuable overviews of WF attack models, but offer limited analysis of several critical factors that fundamentally influence model design and evaluation, including concept drift, single-tab versus multi-tab browsing scenarios, and varying adversarial capabilities. Furthermore, although existing surveys discuss a broad range of defense strategies~\cite{aminuddin2023rise, chao2024systematic, panka2026website}, their classification schemes remain relatively coarse-grained and lack a unified taxonomy that systematically captures the underlying design principles, threat assumptions, deployment requirements, and performance trade-offs of existing approaches. As a result, researchers lack a structured framework for comparing competing techniques and identifying their strengths, limitations, and open research challenges. Another underexplored aspect is the role of datasets in WF research. While recent surveys discuss dataset characteristics and experimental assumptions~\cite{aminuddin2023rise, panka2026website}, they provide limited analysis of dataset diversity and representativeness. Since modern WF attacks and defenses rely heavily on traffic datasets for training and evaluation, dataset selection directly affects the validity, reproducibility, and generalizability of experimental results. The recent survey by ~\citet{panka2026website} expands the discussion to include datasets and experimental evaluation. However, this paper primarily provides a high-level overview of the WF landscape. Consequently, a comprehensive survey that unifies datasets, attacks, and defenses under a common framework, systematically examines their design strategies and assumptions, and highlights emerging challenges and future research opportunities remains absent from the literature. To clearly demonstrate the differences with prior survey papers, \tableautorefname~\ref{tab:relatedwork} compares our survey with existing studies.

\begin{table}
\centering
\scriptsize
\setlength{\tabcolsep}{3pt}
\renewcommand{\arraystretch}{1.2}
\caption{A comparison of related works.}
\label{tab:relatedwork}
\begin{tabular}{|c|c|c|c|c|c|}
\hline
\textbf{Publication} & \textbf{Year} & \textbf{Main Focus} & \textbf{\makecell{Include \\ WF attacks}} & \textbf{\makecell{Include \\ WF defenses}} & \textbf{\makecell{Include \\ Dataset}} \\
\hline
~\citet{karunanayake2021anonymisation} & 2021 & De-Anonymisation Attacks on Tor. &  \LEFTcircle & \Circle & \Circle \\
\hline
~\citet{aminuddin2023rise} & 2023 & Analysis on Tor-based WF techniques and assumptions.  & \CIRCLE & \Circle & \LEFTcircle \\
\hline
~\citet{liu2023survey} & 2023 & Deep learning on Tor-based WF attacks and defenses. & \CIRCLE  & \LEFTcircle & \Circle \\
\hline
~\citet{chao2024systematic} & 2024 & \makecell{Vulnerabilities, Attacks, Defenses, and \\ Formalization in Anonymity Networks.} & \LEFTcircle  & \LEFTcircle & \Circle \\
\hline
~\citet{murad2026website} & 2026 & Overview of representative WF attacks. & \CIRCLE  & \Circle & \Circle \\
\hline
~\citet{panka2026website} & 2026 & \makecell{Overview of WF attacks and defenses, descriptions \\ of datasets, and discussion on experimental evaluation.} & \CIRCLE  & \CIRCLE & \CIRCLE \\
\hline
\textbf{Our work} & \textbf{2026} & \textbf{\makecell{A comprehensive taxonomy and in-depth analysis of the \\ strategies, challenges, and future research directions in \\ Tor-based WF attacks, defenses, and datasets.}} & \CIRCLE & \CIRCLE & \CIRCLE \\
\hline
\multicolumn{6}{l}{\CIRCLE: All prior studies included. \LEFTcircle: Few prior studies included. \Circle: No prior studies included.} \\
\end{tabular}
\vspace{-0.15in}
\end{table}

\subsection{Taxonomies and Our Contributions}
To address these gaps, this survey presents a systematic review of WF research in the Tor network. We develop a unified taxonomy of WF attacks and defenses based on design principles, threat assumptions, feature representations, and operational characteristics. We also provide an in-depth analysis of publicly available datasets, including their collection methodologies, properties, strengths, and limitations. This taxonomy offers a structured perspective on the evolution of WF research and for comparing competing approaches under consistent criteria.

\begin{figure}[t]
  \centering
  \includegraphics[scale=0.30]{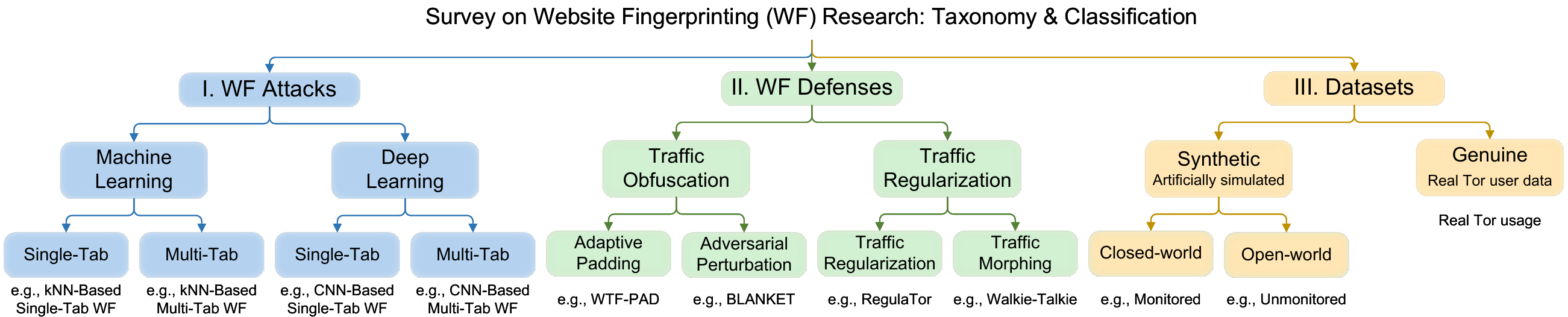}
  \caption{Overview of the taxonomy presented in this survey, categorizing Tor website fingerprinting research into attacks, defenses, and datasets.}
  \Description{Overview of the taxonomy presented in this survey, categorizing Tor website fingerprinting research into attacks, defenses, and datasets.}
  \label{fig:taxonomy}
  \vspace{-0.20in}
\end{figure}

As shown in Fig.~\ref{fig:taxonomy}, WF attacks are categorized according to learning paradigms and browsing scenarios. Specifically, we distinguish between machine learning (ML)-based and deep learning (DL)-based approaches, each further divided into single-tab and multi-tab settings. This organization highlights the transition from handcrafted feature engineering to automated representation learning and emphasizes the challenges introduced by realistic browsing behaviors. For WF defenses, we analyze each approach in terms of its threat model, design objectives, bandwidth and latency overhead, and deployment feasibility. By evaluating defenses under a unified framework, we highlight their strengths and limitations and identify persistent challenges in achieving robust, scalable, and low-overhead protection.

In summary, this work makes the following five major contributions:
\begin{itemize}
    \item \textbf{Unified taxonomy and categorization.} We propose a hierarchical taxonomy from multiple perspectives and organize WF research into datasets, attacks, and defenses.
    \item \textbf{Analysis of challenges and limitations.} We examine fundamental assumptions and limitations of existing methods and identify open research problems.
    \item \textbf{Comprehensive literature review.} We systematically review representative WF studies and analyze their design principles and assumptions.
    \item \textbf{Future research directions.} We identify emerging trends and promising research opportunities for improving WF robustness and practicality.
    \item \textbf{Resources and benchmarks.} We compile publicly available datasets and implementations to support reproducibility and benchmarking.
\end{itemize}

\subsection{Organization of this Survey Paper}

Section~\ref{sec:background} presents the background of Tor, including its architecture, the traffic features used in WF attacks and defenses, evaluation metrics, and the problem formulation of WF research. Section~\ref{sec:review} provides a comprehensive survey of WF datasets, attacks, and defense mechanisms. Section~\ref{sec:challenges} discusses the challenges and limitations of existing WF attacks, defenses, and datasets, and highlights promising future research directions. Finally, Section~\ref{sec:conclusion} concludes the paper.

\section{Background}
\label{sec:background}

This section provides the necessary background for understanding WF attacks and defenses in the Tor network. \sectionautorefname~\ref{subsec:torarc} presents an overview of Tor's architecture and anonymity mechanisms. \sectionautorefname~\ref{subsec:features} reviews the major traffic features that serve as the foundation of Tor traffic analysis. \sectionautorefname~\ref{subsec:metrics} introduces the evaluation metrics and problem formulations commonly used to assess the effectiveness of WF attacks and defenses.

\subsection{Tor Architecture and Anonymity Mechanisms}
\label{subsec:torarc}

Modern privacy-preserving communication systems employ a variety of encryption and anonymity mechanisms to protect user identities and secure data transmission~\cite{dingledine2004tor, bissias2005privacy, clark2013sok}. These mechanisms provide different levels of confidentiality, anonymity, and resistance to traffic analysis, forming the foundation of secure online communications. As summarized in \tablename~\ref{tab:ecryptedtraffic}, Hypertext Transfer Protocol Secure (HTTPS)~\cite{clark2013sok}, Virtual Private Networks (VPNs)~\cite{abbas2023security}, QUIC~\cite{cook2017quic}, and Tor provide different levels of confidentiality, anonymity, and traffic protection. HTTPS employs Transport Layer Security (TLS) to encrypt application-layer data, protecting web content from eavesdropping. VPNs extend protection by encrypting all network traffic between a client and a VPN server; however, users must trust the VPN provider, which can observe both their identity and online activities. QUIC integrates transport-layer encryption over User Datagram Protocol (UDP) to improve performance and reduce latency, but does not provide anonymity. In contrast, Tor combines layered encryption with multi-hop routing to conceal both the content of communication and user identity, offering stronger anonymity guarantees at the cost of increased latency.

\begin{figure}[t]
  \centering
  \includegraphics[scale=0.28]{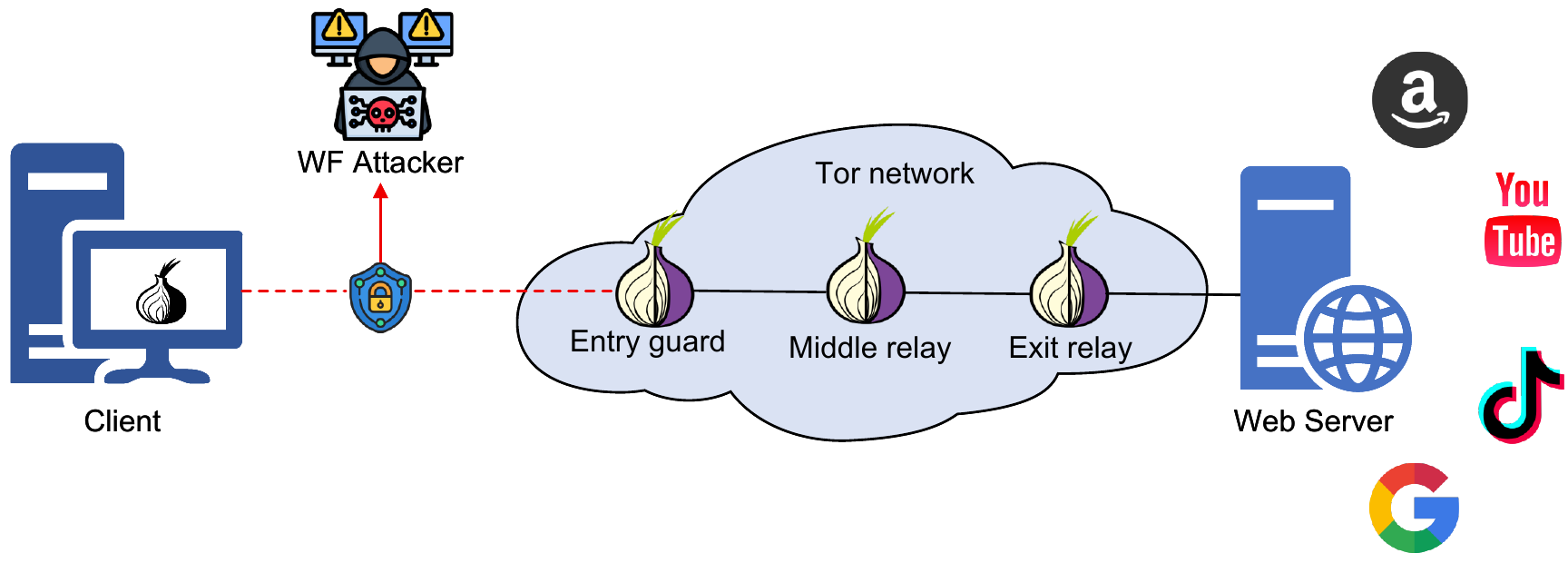}
  \caption{Overview of the Tor network and Threat Model.}
  \Description{Overview of the Tor network and Threat Model.}
  \label{fig:torarch}
  \vspace{-0.10in}
\end{figure}

\begin{table}

\caption{Comparison of representative encrypted communication protocols and anonymity systems.}
\label{tab:ecryptedtraffic}
\scriptsize
\setlength{\tabcolsep}{3pt}
\renewcommand\arraystretch{2}
\centering
\begin{tabular}{|c|c|c|c|c|}
\hline
\textbf{Feature/Layer} & \textbf{HTTPS} & \textbf{VPN} & \textbf{\makecell{QUIC}} & \textbf{Tor (Onion Routing)} \\
\hline
\textbf{Encryption Layer} & Application layer (TLS) & \makecell{Network/Transport layer \\ (\eg, OpenVPN, WireGuard)} & \makecell{Transport layer \\ (over UDP)} & \makecell{Multi-hop onion \\ encryption} \\
\hline
\textbf{Encryption Scope} & \makecell{Application-layer payload \\ (\eg, web content)} & Tunnel traffic & \makecell{Transport-layer \\ payloads} & \makecell{Layered traffic \\ encryption} \\
\hline
\textbf{IP Address Hidden?} & \textcolor{red}{\xmark} No & \makecell{\textcolor{green}{\cmark} Yes from destination \\ \textcolor{red}{\xmark} No from VPN provider} & \textcolor{red}{\xmark} No & \textcolor{green}{\cmark} Yes from destination \\
\hline
\textbf{Anonymity} & \textcolor{red}{\xmark} No & \makecell{Partial (VPN sees \\ identity and activity)} & \textcolor{red}{\xmark} No & \textcolor{green}{\cmark} Yes \\
\hline
\textbf{Traffic Obfuscation} & \textcolor{red}{\xmark} No & Depends on protocol used & \textcolor{red}{\xmark} No & \textcolor{green}{\cmark} Yes \\
\hline
\textbf{\makecell{DNS Leakage \\ Protection}} & Optional via DoH & \makecell{\textcolor{green}{\cmark} Yes, if DNS goes \\ through VPN}  & Optional via DoH & \makecell{\textcolor{green}{\cmark} Yes, DNS requests \\ resolved within Tor} \\
\hline
\textbf{\makecell{Resistant to \\ Censorship?}} &  Limited & \makecell{Depends on obfuscation \\ layer used} &  Limited &  Strong resistance \\
\hline
\textbf{Packet Size} & \makecell{$\sim$1,300 bytes \\ (TCP + TLS payload)} & \makecell{1,400–1,500 bytes \\ (depends on protocol)} &  \makecell{1,200–1,400 bytes \\ (UDP + QUIC)} &  \makecell{Fixed-size \\ (512-byte cells)} \\
\hline
\textbf{Performance} & Fast & Moderate latency increase & Low latency & High latency due to relays \\
\hline
\multicolumn{5}{l}{SNI: Server Name Indication; DoH: DNS over HTTPS; \textcolor{green}{\cmark} Support; \textcolor{red}{\xmark} Not support.} \\
\end{tabular}
\vspace{-0.20in}
\end{table}

Tor achieves anonymity through onion routing, in which traffic is relayed through a sequence of intermediary nodes known as relays~\cite{dingledine2004tor, mccoy2008shining, snader2008tune}. As illustrated in \figurename~\ref{fig:torarch}, a Tor client establishes a circuit consisting of three relays: an entry guard, a middle relay, and an exit relay. A circuit is typically maintained for approximately ten minutes before a new circuit is constructed~\cite{torpathselection}. The entry guard is the only relay aware of the client's IP address, while the exit relay forwards traffic to the destination server. The middle relay serves as an additional layer of indirection, preventing direct linkage between the entry and exit points. To further protect user privacy, traffic is encrypted in multiple layers before transmission. Each relay decrypts only a single layer of encryption and learns information about only its immediate predecessor and successor. Consequently, no individual relay can simultaneously determine both the source and destination of a communication, thereby preserving user anonymity. This architecture forms the foundation of Tor's resistance to surveillance, traffic correlation, and censorship, making it the primary deployment platform studied in contemporary research on website fingerprinting.

Tor also incorporates a layered encapsulation and segmentation mechanism to ensure secure and reliable data transmission. At the application layer, data are divided into fixed-size Tor cells of 512 bytes, which constitute the fundamental communication unit within the network. These cells are encrypted and encapsulated into TLS records, each of which typically contains multiple Tor cells. As TLS records traverse the network stack, they are further segmented into TCP packets, whose sizes are constrained by the maximum segment size (MSS), commonly 1460 bytes in Ethernet networks. Furthermore, a single Tor circuit can multiplex multiple streams, where each stream corresponds to an independent TCP connection. To regulate data transmission, Tor employs a flow-control mechanism based on \textit{SENDME} cells. Specifically, a stream-level \textit{SENDME} cell is generated for every 50 received cells, whereas a circuit-level \textit{SENDME} cell is generated for every 100 received cells. Through the combined effects of cell multiplexing, TLS encapsulation, TCP segmentation, and flow-control signaling, Tor produces complex traffic patterns that reflect the substantial and heterogeneous data exchanges generated by modern web applications~\cite{panchenko2016website,deng2025countmamba}.

\subsection{Tor Traffic Features}
\label{subsec:features}

Traffic features constitute the foundation of WF attacks and defenses. They represent observable characteristics of encrypted network traffic that can be exploited to infer visited websites or, conversely, targeted by defenses to reduce information leakage. Existing WF research utilizes a diverse set of features ranging from fine-grained packet-level representations, such as packet directions~\cite{rahman2019tik, shi2025multiscale, zou2024relation, zhao2024towards}, packet lengths~\cite{wang2014effective, liberatore2006inferring}, timing patterns~\cite{dyer2012peek, wang2014effective, kwon2015circuit, attarian2019adawfpa}, traffic bursts~\cite{wang2014effective, zhao2024towards, bahramali2023realistic}, and packet ordering~\cite{cai2012touching, wang2013improved}, to coarse-grained statistics~\cite{zhu2025wf, deng2025countmamba}, including traffic volume, bandwidth consumption, and trace duration. While attack models leverage these features to distinguish websites, defense mechanisms seek to obfuscate, perturb, or reshape them to reduce fingerprintability. As summarized in \tablename~\ref{tab:torfeatures}, the WF literature can be broadly organized around six major feature categories: traffic timing, traffic length, packet ordering, traffic bursts, traffic direction, and coarse-grained traffic statistics.

\begin{table*}
\centering
\scriptsize
\setlength{\tabcolsep}{3pt}
\renewcommand{\arraystretch}{1.2}
\caption{The summary of important features and how to use the features for WF attack and defense techniques.}
\label{tab:torfeatures}

\begin{tabular}{cp{5cm}l} 
\toprule[0.75pt]
\rowcolor{olive!25}
\textbf{Feature Categories} & \textbf{Representative Uses in WF Attacks} & \textbf{Representative Uses in WF Defenses} \\
\midrule
\multirow{5}{*}{Traffic timing} & Packet inter-arrival times~\cite{wang2016realistically}. &   \\
 & Time gaps~\cite{xu2018multi}. & Random or adaptive dummy packets to vary the timing  \\
 & Total transmission time~\cite{pulls2020towards, dyer2012peek, wang2014effective, kwon2015circuit, attarian2019adawfpa}. & patterns~\cite{shmatikov2006timing, cui2018realistic}. \\ 
& Delay between the first outgoing and incoming & Traffic regularization to normalize the timing patterns~\cite{gong2020zero}. \\
&  packet~\cite{xu2018multi, yin2021automated}. & \\
\rowcolor{gray!25}
 & Total number of packets~\cite{wang2016realistically, wang2014effective, liberatore2006inferring}. &  \\
 \rowcolor{gray!25}
 & Number of outgoing packets~\cite{wang2013improved}. &  \\
 \rowcolor{gray!25}
\multirow{-3}{*}{Packet length} & Number of incoming packets~\cite{wang2013improved}. & \multirow{-3}{*}{\makecell[l]{Dummy packets will obfuscate the packet length.}}\\

\multirow{4}{*}{Packet ordering} & Outgoing and incoming packet sequen- & Introduce randomness through dummy tr-  \\ 
 & ces (direction and timing)~\cite{cai2012touching, wang2013improved, zhao2024towards, zhou2023wf}. & affic and artificial delays~\cite{abusnaina2020dfd, kadianakis2021tor, dyer2012peek, cai2014cs}. \\
 & Statistical properties of packet-order se- & Traffic morphing~\cite{chan2018website, al2019bimorphing, shen2024real} and creating a \\
 & quence (direction and timing)~\cite{attarian2019adawfpa, wang2013improved, wang2016realistically}. & single "supersequence" of packets~\cite{wang2014effective}.\\
 \rowcolor{gray!25}
  & Burst-frequency distributions~\cite{wang2014effective, zhao2024towards}. & Injecting dummy packets~\cite{abusnaina2020dfd, kadianakis2021tor}.\\
 \rowcolor{gray!25}
  & Burst length and burst size statistics~\cite{bahramali2023realistic}. & Traffic regularization~\cite{dyer2012peek, gong2020zero, holland2020regulator}.\\
 \rowcolor{gray!25}
\multirow{-3}{*}{Traffic bursts}  & Sequences of packet bursts~\cite{panchenko2016website, hayes2016k}. & Using adversarial perturbation~\cite{shen2024real, li2022minipatch, gong2022surakav}. \\

\multirow{2}{*}{Direction} & Assigning +1 to outgoing packets and & \multirow{2}{*}{Defense models cannot change the packet direction.} \\
 & -1 to incoming packets~\cite{rahman2019tik, shi2025multiscale, zou2024relation, zhao2024towards}. &  \\

\rowcolor{gray!25}
 & Bandwidth consumption~\cite{dyer2012peek}. & Regulate transmission rates to flatten bandwidth variations.  \\
 \rowcolor{gray!25}
 &  Traffic "burstiness" and burst bandwidth~\cite{zhu2025wf}. & Morph aggregate traffic distributions to resemble other web- \\
\rowcolor{gray!25}
\multirow{-3}{*}{Coarse features} &  Time-window-based counting~\cite{deng2025countmamba}. & sites. \\
\bottomrule[0.75pt]
\end{tabular}
\vspace{-0.20in}
\end{table*}

\subsubsection{Traffic Timing}
\label{subsubsec:timing}

Traffic timing features capture the temporal characteristics of encrypted traffic and have long been recognized as an important source of information leakage in website fingerprinting. Common timing-related features include packet inter-arrival times~\cite{wang2016realistically}, round-trip times (RTTs)~\cite{jansen2026cellshift}, trace durations, and idle periods within a traffic trace~\cite{xu2018multi, yin2021automated}. Additional metrics, such as the ratio of active transmission time to total trace duration~\cite{dyer2012peek, wang2014effective, kwon2015circuit, attarian2019adawfpa}, characterize the temporal distribution of communication activity. Collectively, these features capture website-specific loading behaviors and have been shown to improve classification performance in both traditional ML- and DL-based WF attacks~\cite{wang2016realistically, dyer2012peek, panchenko2011website, wang2013improved, mei2025high, gao2024multi}.

Timing features present a challenging surface to protect. Unlike packet-ordering and traffic bursts, which can be easily padded or reshaped, time-based features are inherently tied to user interaction, network conditions, and server response behavior. Defenses such as adaptive padding~\cite{witwer2022padding, juarez2016toward}, timing obfuscation~\cite{shmatikov2006timing, lu2023lightweight}, and randomized delays~\cite{hong2022website} have been proposed to mitigate timing-based attacks. However, these approaches often introduce latency or degrade performance. As a result, incorporating traffic-timing awareness into both attacks and defenses is now essential for evaluating the real-world feasibility and robustness of WF techniques.

\subsubsection{Packet Length}
\label{subsubsec:length}

Packet length features characterize the overall volume of a traffic trace and are typically measured using packet counts, cell counts, byte counts, or trace duration~\cite{wang2014effective, hayes2016k}. Since websites vary considerably in their content composition, page complexity, and resource demands, the volume of generated traffic often differs substantially across websites~\cite{wang2013improved, liberatore2006inferring, wang2016realistically}. However, accurate measurement of traffic length typically requires precise identification of traffic trace boundaries, which is challenging in continuous network traffic. Moreover, users often browse multiple webpages simultaneously, generating overlapping traffic streams that make it difficult to attribute packets to individual webpage visits. To address these challenges, several studies employ time-gap-based approximations to estimate traffic boundaries~\cite{attarian2019adawfpa, yin2021automated, xu2018multi}. Despite its simplicity, packet length continues to provide complementary information when combined with more discriminative feature categories.

\subsubsection{Packet Ordering}
\label{subsubsec:ordering}

Packet ordering features describe the sequential arrangement of packets or cells within a traffic trace. These features capture request-response dependencies and webpage-loading behaviors that remain observable even with encryption~\cite{cai2012touching, wang2013improved}. Prior work has shown that ordered packet sequences often contain website-specific structural patterns that can significantly improve classification performance when combined with timing, directional, and burst-level information~\cite{wang2013improved, zhao2024towards, zhou2023wf}.

For defense mechanisms, obfuscating packet ordering is inherently difficult because packet sequences are closely tied to application functionality and network performance. Significant modifications to packet ordering may disrupt communication semantics, increase latency, and degrade user experience. Existing defenses therefore rely on indirect mechanisms, such as adaptive packet injection~\cite{abusnaina2020dfd}, traffic regularization~\cite{kadianakis2021tor, chan2018website, holland2020regulator}, and traffic morphing~\cite{wang2017walkie, al2019bimorphing}, to reduce the stability of ordering patterns.

\subsubsection{Traffic Bursts}
\label{subsubsec:bursts}

Traffic bursts represent sequences of consecutive packets traveling in the same direction and constitute one of the most informative feature classes in contemporary WF research. Burst-level features, including burst size, burst count, and burst ordering, capture higher-level communication structures that remain relatively stable across repeated visits to the same website~\cite{bahramali2023realistic}.

Due to their strong discriminative power, burst representations are extensively employed in WF attacks and are a primary target of modern defense mechanisms~\cite{wang2026stmwf, wang2026cease}. Representative defenses, such as Walkie-Talkie~\cite{wang2017walkie} and WTF-PAD~\cite{juarez2016toward}, explicitly aim to reshape or conceal burst characteristics. Nevertheless, preserving usability while effectively obfuscating burst structures remains a significant challenge.

\subsubsection{Traffic Direction}
\label{subsubsec:direction}

Traffic direction features represent the sequence of incoming and outgoing packets within a traffic trace and are commonly encoded as directional sequences. These features capture the communication structure between browsers and web servers and are widely regarded as one of the most robust sources of information leakage in WF attacks~\cite{rahman2019tik, shi2025multiscale, zou2024relation, zhao2024towards}. Unlike timing or packet burst features, traffic direction cannot be modified. Although defense mechanisms may partially obscure directional patterns through traffic obfuscation and traffic regularization, direction-based representations continue to play a central role in WF attack models.

\subsubsection{Coarse-grained Features}

Coarse-grained features describe aggregate traffic characteristics rather than packet-level behavior. Representative examples include total bandwidth consumption, per-direction bandwidth, trace duration, and traffic burstiness~\cite{dyer2012peek, deng2025countmamba, zhu2025wf, tan2024inter}. Although these features provide less detailed information than fine-grained packet representations, they often remain effective even when lower-level features are partially obfuscated.

Compared with packet-level feature categories, coarse-grained features have received relatively limited attention in recent WF research. Nevertheless, existing studies suggest that aggregate traffic statistics can still reveal significant information about visited websites, highlighting the need for future defense mechanisms that explicitly address both fine-grained and coarse-grained sources of information leakage.

\subsection{Evaluation Metrics and Problem Formulation}
\label{subsec:metrics}


In a typical WF scenario, the traffic exchanged between a Tor client and its entry relay is observed as a time-ordered sequence of packets. Formally, a traffic trace associated with loading a webpage can be represented as
$w = [(t_{1}, d_{1}), (t_{2}, d_{2}), \ldots, (t_{N}, d_{N})]$,
where $N$ denotes the total number of packets observed during the session, $t_i$ represents the timestamp of the $i$-th packet, and $d_{i}$ denotes its direction, for $1 < i\le N$. Since Tor encapsulates application-layer data into fixed-size cells, packet size provides limited discriminative information for WF attacks. Consequently, the direction variable is typically defined as $d_i \in \{+1,-1\}$, where $+1$ and $-1$ indicate outgoing and incoming packets, respectively~\cite{herrmann2009website, panchenko2011website, cai2012touching, wang2013improved}. This representation captures both temporal and directional characteristics of network traffic, which are the primary features exploited by modern WF attacks to distinguish webpages despite encryption and traffic padding.

\subsubsection{Problem Formulation}
\emph{\textbf{Data Drift:}}
In ML and DL, data drift (also known as covariate shift) occurs when the statistical distribution of input data changes over time, creating a mismatch between the environment in which a model was trained and the live environment where it is deployed. Because models are optimized based on the patterns of the training set, this divergence often leads to ``model decay", where predictive performance and reliability degrade significantly~\cite{moreno2012unifying}. Importantly, under data drift, the underlying relationship between inputs and labels remains unchanged.

A mathematical representation of data drift can be expressed as follows:
\begin{equation}
    p_{train}(X) \ne  p_{test}(X),\ p(y\mid X)\ \text{unchanged}
\end{equation}
Where $p(X)$ denotes the probability distribution of input features, and $p(y\mid X)$ is the conditional probability of the target given the features (the model's "logic").

\emph{\textbf{Concept Drift:}}
In contrast to data drift, concept drift specifically refers to a shift in the underlying relationship between input features and the target variable. In these scenarios, the statistical properties of the target variable itself change, meaning that even if the input data remains consistent, the model's predictions become increasingly inaccurate because the ``ground truth" logic it once learned is no longer valid~\cite{gama2014survey, shusterman2025understanding}. 

Formally, concept drift between time points $t_{0}$ and time point $t_{1}$ can be defined as:
\begin{equation}
    \exists X: p_{t_{0}}(X, y) \ne  p_{t_{1}}(X, y)
\end{equation}
where $p_{t_{0}}(X, y)$ denotes the joint distribution of the input variables $X$ and the target variable $y$ at time $t_{0}$.Such changes can be characterized by variations in different components of this distribution: (1) shifts in the class prior probabilities $p(y)$; 2) changes in the class-conditional distributions $p(X\mid y)$; and consequently, 3) changes in the posterior probabilities $p(y\mid X)$, which directly affect model predictions.

\subsubsection{Attack Performance Metrics}
WF attack is a traffic analysis technique that aims to infer the website, webpage, or online service accessed by a user based solely on observable network traffic characteristics, even when encryption is in use. Rather than inspecting packet contents, a WF attacker exploits side-channel information exposed by encrypted traffic, such as packet directions, timings, burst patterns, and flow dynamics~\cite{sun2002statistical, dyer2012peek, panchenko2016website, yan2018feature}. Because different webpages generate distinct sequences of network activities during loading and interaction, these traffic characteristics can serve as unique fingerprints that enable website identification~\cite{bissias2005privacy, liberatore2006inferring, lu2010website, guo2021website}.

Formally, let $\mathcal{W}=\{w_{1},w_{2},\ldots,w_{m}\}$ denote a set of observed traffic traces and $\mathcal{C}={c_{1},c_{2},\ldots,c_{m}}$ denote the corresponding website labels. The objective of a WF attack is to learn a classification function $f: \mathcal{W} \rightarrow \mathcal{C}$, where $\mathcal{W}$ represents the space of network traffic traces and $f(w)$ predicts the website associated with an observed trace $w$. Given a previously unseen traffic trace, the learned classifier aims to infer the corresponding website based solely on observable traffic characteristics, despite the underlying communication being encrypted. Modern WF attacks predominantly employ machine learning and deep learning techniques to automatically learn discriminative representations from traffic features, thereby improving classification accuracy and robustness under diverse network conditions.

Evaluating the effectiveness of WF attacks requires quantitative metrics that measure an attacker's ability to identify visited websites from encrypted traffic traces correctly. The choice of evaluation metric depends on the underlying threat model and experimental setting. In closed-world scenarios, where all traffic traces belong to a predefined set of monitored websites, classification accuracy is commonly used to assess attack performance. Open-world scenarios introduce a large number of unmonitored websites, requiring additional metrics that capture the trade-off between correctly identifying monitored websites and avoiding false alarms.

\emph{\textbf{Accuracy / Attack Success Rate (ASR):}} In closed-world settings, accuracy is the most widely used metric for evaluating WF attacks. It measures the proportion of traffic traces that are correctly classified by an attacker $Accuracy = ASR = \frac{N_{acc}}{N_{all}}$, where $N_{acc}$ denotes the number of correctly classified traffic traces and $N_{all}$ denotes the total number of evaluated traces. A higher accuracy indicates a more effective attack model.

\emph{\textbf{TPR and FPR:}}
Open-world evaluations commonly employ TPR and FPR to characterize an attacker's ability to distinguish monitored websites from unmonitored websites. TPR measures the proportion of monitored traces correctly identified as monitored, whereas FPR measures the proportion of unmonitored traces incorrectly classified as monitored. These metrics are defined as $TPR = \frac{TP}{TP + FN}$ and $FPR = \frac{FP}{FP + TN}$, where $TP$ and $TN$ denote the numbers of correctly classified monitored and unmonitored traces, respectively, and $FP$ and $FN$ denote the numbers of incorrectly classified monitored and unmonitored traces. An effective WF attack generally achieves a high TPR while maintaining a low FPR.

\emph{\textbf{Precision, Recall, and F1-Score.}}
Precision, recall, and the F1-score are widely applied in open-world evaluations, where a substantial class imbalance exists between monitored and unmonitored websites. Precision measures the proportion of traces predicted as monitored that are actually monitored, while recall measures the proportion of monitored traces that are correctly identified and is equivalent to the true positive rate (TPR). These metrics are defined as: $Precision = \frac{TP}{TP + FP}$, $Recall = \frac{TP}{TP + FN}$, and $F1 = \frac{2 \times Precision \times Recall}{Precision + Recall}=\frac{2TP}{2TP + FP + FN}$. The F1-score provides a single measure that balances precision and recall through their harmonic mean, making it particularly useful in imbalanced classification settings. A higher F1-score indicates a more favorable trade-off between correctly identifying monitored websites and minimizing false alarms. In addition, Precision–Recall (PR) curves are frequently used to visualize the relationship between precision and recall across different classification thresholds, providing a more informative assessment of classifier performance in open-world scenarios where class distributions are highly skewed.

\subsubsection{Defense Overhead Metrics}
The objective of a WF defense is to reduce the amount of information that an adversary can extract from encrypted traffic traces while preserving balanced network performance. Existing defenses~\cite{wright2009traffic, nithyanand2014glove, panchenko2016website} achieve this goal by obfuscating observable traffic characteristics, such as packet timing patterns and burst structures, thereby reducing the distinguishability of traffic traces associated with different websites.

Despite significant advances, existing WF defenses exhibit several important limitations. First, many defenses remain vulnerable to adaptive adversaries that explicitly train attack models on defended traffic traces~\cite{khajavi2026lightening, shen2024real, qiao2024trace}. Second, defense effectiveness is typically evaluated under specific datasets, network conditions, and threat models~\cite{xie2025gapdis, li2022minipatch, rahman2020mockingbird}, raising concerns about the ability of these defenses to generalize across diverse browsing environments, evolving website content, and real-world deployment scenarios. Third, most defenses rely on deterministic traffic transformations~\cite{juarez2016toward, holland2020regulator, gong2020zero}, which may allow adversaries to learn or approximate the underlying transformation process and recover residual identifying information from defended traces.

From a formal perspective, a WF defense can be modeled as a transformation from an original traffic trace to a defended traffic trace:
\begin{equation}
f: \mathcal{W}\rightarrow \mathcal{W}^{'}, \quad w \mapsto w^{'},
\end{equation}
where $\mathcal{W}$ denotes the space of original traffic traces and $\mathcal{W}^{'}$ denotes the space of defended traces. Under this formulation, existing defenses can be broadly categorized according to the mapping relationship induced by the transformation function.

Many traffic-obfuscation approaches, including FRONT~\cite{gong2020zero}, TrafficSliver~\cite{de2020trafficsliver}, and Minipatch~\cite{li2022minipatch}, can be viewed as one-to-one mappings, where each original trace is mapped to a unique defended trace:
\begin{equation}
\forall w \in \mathcal{W},\ \exists ! w' \in \mathcal{W}^{'},\quad f(w)=w^{'}.
\end{equation}
In contrast, traffic-regularization approaches, such as Regulator~\cite{holland2020regulator} and Palette~\cite{shen2024real}, exhibit many-to-one behavior by transforming multiple traffic traces into a common traffic representation:
\begin{equation}
\exists w_1,w_2, \dots \in W,\ w' \in W',\quad
f(w_1)=f(w_2)=f(\dots)=w'
\Rightarrow w_1\neq w_2\neq \dots.
\end{equation}

\emph{\textbf{Bandwidth Overhead.}}
Bandwidth overhead quantifies the additional traffic introduced by a defense mechanism, typically through dummy packet injection, padding, or traffic morphing. It is defined as $B(w,w') = \frac{|w'|-|w|}{|w|}$, where $w$ and $w'$ denote the original and defended traffic traces, respectively, and $|w|$ and $|w'|$ represent their corresponding traffic volumes. Lower bandwidth overhead indicates a more communication-efficient defense.

\emph{\textbf{Latency Overhead.}}
Latency overhead measures the additional transmission delay introduced by a defense mechanism. It is defined as $T(w,w') = \frac{t' - t}{t}$, where $t$ and $t'$ denote the completion times of the original and defended traffic traces, respectively. Lower latency overhead generally corresponds to a better user experience and improved deployability in real-world environments.

\section{Survey on WF Research}
\label{sec:review}

This section presents a systematic classification of WF research, encompassing attacks, defenses, and datasets. WF attacks are categorized into traditional machine learning and deep learning approaches, with additional discussion of single-tab and multi-tab attack scenarios. WF defenses are classified into adaptive padding, adversarial perturbation, traffic regularization, and traffic morphing techniques. We also provide an overview of commonly used datasets that support the development and evaluation of WF attacks and defenses.

\subsection{WF Attacks}
\label{subsec:bacwfa}

WF attacks are a class of traffic analysis techniques that aim to infer a user's browsing activity by analyzing observable characteristics of encrypted network traffic~\cite{panchenko2016website, wang2013improved, hayes2016k, aminuddin2023rise, fu2024detecting}. Unlike attacks that target encrypted content directly, WF attacks exploit side-channel information, such as packet directions, timing patterns, and traffic bursts, to identify visited websites~\cite{shusterman2025understanding}. Consequently, WF attacks pose a significant threat to anonymity systems, particularly Tor, where communication contents are encrypted, but traffic metadata remains observable.

The evolution of WF attacks can be broadly divided into two stages: traditional ML-based approaches and DL-based approaches. Traditional methods rely heavily on handcrafted traffic features and classical classification algorithms, whereas modern DL approaches learn discriminative representations directly from raw traffic traces, substantially improving attack effectiveness and generalization capabilities. \tableautorefname~\ref{tab:wfa} presents a comprehensive overview of representative WF attack models, the traffic features they exploit, and the browsing scenarios under which they have been evaluated. 

Early WF research primarily targeted encrypted communication protocols such as HTTPS~\cite{ sun2002statistical, bissias2005privacy}, OpenSSH~\cite{hintz2002fingerprinting}, and OpenVPN~\cite{liberatore2006inferring, lu2010website}. These studies demonstrated that traffic characteristics alone could reveal substantial information about user activities. The first WF attack against Tor was proposed by ~\citet{herrmann2009website}, who employed a Na\"ive Bayes classifier using traffic direction and traffic bursts as features. Although the reported accuracy was limited, this work established the feasibility of fingerprinting Tor traffic and motivated extensive subsequent research.

\begin{table}[t]
\centering
\scriptsize
\setlength{\tabcolsep}{3pt}
\renewcommand{\arraystretch}{1.2}
 \caption{Summary of representative website fingerprinting attack models, targeted browsing scenarios, and exploited traffic features.}
\label{tab:wfa}
\begin{tabular}{|c|c|c|c|c|c|c|c|c|c|c|c|}
\hline
\multicolumn{3}{|c|}{\textbf{Model}} & \textbf{\makecell{Traffic \\ Timing}} & \textbf{\makecell{Circuit \\ Length}} & \textbf{\makecell{Traffic \\ Direction}} & \textbf{\makecell{Packet \\ Ordering}} & \textbf{\makecell{Traffic \\ Bursts}} & \textbf{\makecell{Coarse \\ Features}} & \textbf{\makecell{Single-tab \\ Studies}}  & \textbf{\makecell{Multi-tab \\ Studies}} \\
\hline
\multirow{43}{*}{\rotatebox[origin=c]{90}{Machine Learning}} & \multicolumn{2}{c|}{\multirow{4}{*}{Na\"ive Bayes}} & \xmark & \xmark & \cmark & \xmark & \cmark & \xmark & ~\cite{herrmann2009website} & \textbf{-} \\
\cline{4-11}
& \multicolumn{2}{c|}{} & \xmark & \cmark & \cmark & \xmark & \xmark & \xmark & \cite{liberatore2006inferring} & \textbf{-} \\
\cline{4-11}
& \multicolumn{2}{c|}{} & \cmark & \cmark & \cmark & \cmark & \xmark & \xmark &~\cite{wang2016realistically} & \textbf{-} \\
\cline{4-11}
& \multicolumn{2}{c|}{} & \cmark & \xmark & \cmark & \xmark & \cmark & \cmark & ~\cite{dyer2012peek} & \textbf{-} \\
\cline{2-11}
& \multicolumn{2}{c|}{\multirow{4}{*}{SVM}} & \xmark & \xmark & \cmark & \cmark & \xmark & \xmark &~\cite{cai2012touching, wang2013improved} & \textbf{-} \\
\cline{4-11}
& \multicolumn{2}{c|}{} & \cmark & \cmark & \cmark & \cmark & \xmark & \xmark &~\cite{wang2016realistically} & \textbf{-} \\
\cline{4-11}
& \multicolumn{2}{c|}{} & \cmark & \cmark & \cmark & \xmark & \xmark & \xmark & \cite{panchenko2011website} & \textbf{-} \\
\cline{2-11}
&  \multicolumn{2}{c|}{\multirow{6}{*}{k-NN}} & \cmark & \cmark & \cmark & \cmark & \cmark & \xmark &~\cite{wang2014effective} & \textbf{-} \\
\cline{4-11} 
& \multicolumn{2}{c|}{} & \cmark & \xmark & \cmark & \xmark  & \cmark & \xmark & \textbf{-} &~\cite{zhao2024towards} \\
\cline{4-11}
& \multicolumn{2}{c|}{} & \xmark &  \xmark & \cmark & \cmark & \xmark & \xmark &~\cite{wang2013improved} & \textbf{-} \\
\cline{4-11}
& \multicolumn{2}{c|}{} & \cmark & \cmark & \cmark & \cmark & \xmark & \xmark &~\cite{wang2016realistically} & \textbf{-} \\
\cline{4-11}
& \multicolumn{2}{c|}{} & \xmark & \xmark & \cmark & \xmark & \xmark &  \xmark &~\cite{zhuo2017website} & \textbf{-} \\
\cline{4-11}
& \multicolumn{2}{c|}{} & \cmark & \cmark & \cmark  &  \xmark & \xmark & \xmark &  \textbf{-} &~\cite{cui2019revisiting} \\
\cline{2-11}
& \multicolumn{2}{c|}{Decision Tree}  & \cmark & \xmark & \cmark & \cmark & \cmark & \xmark & ~\cite{kwon2015circuit} & \textbf{-} \\
\cline{2-11}
& \multicolumn{2}{c|}{Random Forest} & \cmark & \cmark & \cmark & \cmark & \cmark & \xmark &~\cite{hayes2016k} & \textbf{-} \\
\cline{2-11}
& \multicolumn{2}{c|}{Hoeffding Tree} & \cmark & \cmark & \cmark & \cmark & \cmark & \xmark &~\cite{attarian2019adawfpa} & \textbf{-} \\
\cline{2-11}
& \multicolumn{2}{c|}{\multirow{2}{*}{\makecell{BalanceCascade-\\XGBoost}}} & \cmark & \cmark & \cmark & \cmark & \cmark & \xmark & \textbf{-} & ~\cite{xu2018multi} \\
\cline{4-11}
& \multicolumn{2}{c|}{} & \cmark & \cmark & \cmark & \xmark & \cmark & \xmark & \textbf{-} & ~\cite{yin2021automated} \\
\cline{2-11}
& \multirow{25}{*}{\rotatebox[origin=c]{90}{Deep Learning}} & \multirow{7}{*}{CNN} & \cmark & \xmark & \cmark & \xmark & \xmark & \xmark &~\cite{hong2024website, shen2023subverting, bhat2018var, gong2024wfcat, cui2023edf} & ~\cite{yuan2025mw3f, cui2020more, wang2026stmwf}\\
\cline{4-11}
& & & \cmark & \xmark & \cmark & \xmark & \cmark & \xmark &~\cite{rahman2019tik} & ~\cite{zhao2024towards}\\
\cline{4-11}
& & & \cmark & \xmark  & \cmark & \xmark & \xmark & \xmark &~\cite{rimmer2017automated,   wang20202ch, mei2025high} & ~\cite{guan2021bapm, yin2024traces, deng2026enhancing} \\
\cline{4-11}
& & &\cmark & \cmark & \cmark & \xmark & \cmark & \xmark & \cite{yuan2025class, jeon2026rofire} & ~\cite{deng2024robust, cui2019revisiting, huang2026forge} \\
\cline{4-11}
& & & \xmark &  \xmark & \cmark & \xmark & \xmark & \xmark &~\cite{wang2022snwf, sirinam2018deep, pulls2020website,  chen2022tform, qian2024enhancing, ding2024multi} & ~\cite{deng2023robust, deng2026towards} \\
\cline{4-11}
& & & \xmark & \xmark & \cmark & \xmark & \cmark & \xmark &~\cite{tian2023tiny} & \textbf{-} \\
\cline{4-11}
&  & & \cmark & \xmark & \cmark & \cmark & \xmark & \cmark & \textbf{-} & ~\cite{deng2025countmamba} \\
\cline{3-11}
& & LSTM\textbackslash SADE & \cmark & \xmark & \cmark & \xmark & \xmark & \xmark &~\cite{rimmer2017automated} & \textbf{-} \\
\cline{3-11}
& & \multirow{5}{*}{\makecell{Few-shot \\ Learning}} & \cmark & \xmark & \cmark & \xmark & \xmark & \xmark &~\cite{hong2024website,  shen2025swallow} & ~\cite{meng2025beyond} \\
\cline{4-11}
& & & \xmark & \xmark & \cmark & \xmark & \xmark & \xmark &~\cite{ tan2024adaptability, wang2026data} & \\
\cline{4-11}
& & &  \xmark & \cmark & \cmark &  \xmark & \xmark & \xmark &~\cite{sirinam2019triplet, wang2024wf3a} & \textbf{-} \\
\cline{4-11}
& & & \xmark & \xmark & \cmark & \xmark & \cmark & \xmark &~\cite{xie2024contrastive, li2025cross} & \textbf{-} \\
\cline{4-11}
& & &  \cmark & \xmark & \cmark & \xmark &  \cmark &  \cmark &~\cite{zhu2025wf} & \textbf{-} \\
\cline{3-11}
& & \multirow{2}{*}{DNN} &  \cmark  & \cmark & \cmark & \xmark & \xmark & \xmark&~\cite{oh2017p} & \textbf{-} \\
\cline{4-11}
& & & \cmark & \xmark & \cmark & \xmark & \xmark & \xmark &~\cite{singh2023first} & \textbf{-} \\
\cline{3-11}
& & \multirow{2}{*}{GAN} & \cmark & \cmark & \cmark & \xmark & \xmark & \xmark &~\cite{oh2021gandalf} & \textbf{-} \\
\cline{4-11}
& & & \xmark & \xmark & \cmark & \xmark & \xmark & \xmark & \textbf{-} & ~\cite{xie2023multi} \\
\cline{3-11}
& & \multirow{3}{*}{Transformer} & \cmark & \xmark & \cmark & \xmark & \xmark & \xmark & \textbf{-} & ~\cite{jin2023transformer} \\
\cline{4-11}
& & & \cmark & \xmark & \cmark & \cmark & \xmark & \xmark & \textbf{-} &~\cite{mathews2024laserbeak} \\ 
\cline{4-11}
& & & \xmark & \xmark & \cmark & \cmark & \xmark & \xmark &~\cite{zhou2023wf} & \textbf{-} \\
\cline{3-11}
& & \multirow{2}{*}{GNN} & \cmark & \cmark & \cmark & \xmark & \xmark &  \xmark &~\cite{gao2024multi} & \textbf{-} \\
\cline{4-11}
& & & \xmark & \xmark & \cmark & \xmark & \xmark & \cmark &~\cite{tan2024inter} & \textbf{-} \\
\cline{3-11}
& & SimCLR & \xmark & \xmark & \cmark & \xmark & \cmark & \xmark &~\cite{bahramali2023realistic} & \textbf{-} \\
\cline{3-11}
& & LLMs & \cmark & \xmark & \cmark & \xmark & \cmark & \xmark &~\cite{zhao2026robust} & \textbf{-} \\
\hline
\end{tabular}
\vspace{-0.20in}
\end{table}

\subsubsection{Traditional Machine Learning-Based Attacks}

Traditional WF attacks typically employ manually engineered features derived from packet directions, packet lengths, timing information, and burst statistics. These features are subsequently processed by supervised learning algorithms to classify websites. Representative classifiers include Na\"ive Bayes~\cite{herrmann2009website}, Support Vector Machines (SVMs)~\cite{panchenko2011website, cai2012touching, wang2013improved}, k-Nearest Neighbors (k-NN)~\cite{wang2014effective}, decision trees~\cite{kwon2015circuit}, and random forests~\cite{hayes2016k}. Among these approaches, SVM-based and k-NN-based attacks achieved particularly strong performance in early Tor fingerprinting studies. ~\citet{panchenko2011website} demonstrated that carefully designed traffic features combined with SVM classifiers could significantly outperform earlier statistical methods. Subsequently, ~\citet{wang2014effective} introduced the k-fingerprinting attack, which leveraged random forests and nearest-neighbor techniques to achieve high classification accuracy in both closed-world and open-world settings. Despite their effectiveness, traditional ML approaches depend heavily on feature engineering and often exhibit limited robustness when network conditions, browsing behaviors, or website content evolve.

\subsubsection{Deep Learning-Based Attacks}

Recent advances in deep learning have fundamentally transformed WF research. Unlike traditional approaches that require manually designed features, DL-based attacks automatically learn hierarchical traffic representations directly from raw packet sequences and traffic traces. This capability enables the extraction of complex spatial, temporal, and structural patterns that are difficult to capture using handcrafted features.

Representative DL architectures employed in WF attacks include Convolutional Neural Networks (CNNs)~\cite{abe2016fingerprinting, sirinam2018deep, rimmer2017automated}, Long Short-Term Memory (LSTM) networks~\cite{rimmer2017automated}, Generative Adversarial Networks (GANs)~\cite{oh2021gandalf}, Transformer-based models~\cite{zhou2023wf, mathews2024laserbeak}, and Graph Neural Networks (GNNs). Among these approaches, CNN-based models such as Deep Fingerprinting (DF)~\cite{sirinam2018deep} demonstrated that deep architectures could substantially outperform traditional ML methods, achieving classification accuracies exceeding 90\% in many evaluation settings. Subsequent research has further improved performance through attention mechanisms, self-supervised learning, multi-modal feature fusion, and large-scale Transformer architectures.

Despite their impressive performance, deep learning-based WF attacks typically require large-scale labeled datasets for training. Collecting such datasets is both time-consuming and resource-intensive, as it involves repeatedly crawling websites under diverse network conditions and continuously updating traffic traces to account for content changes over time~\cite{sirinam2019triplet, wang2024wf3a}. To address this challenge, recent research has explored few-shot learning techniques for WF attacks~\cite{zou2022efficient, hong2024website, meng2025beyond, shen2025swallow}. Few-shot learning aims to learn effective traffic representations from only a small number of labeled examples, thereby reducing the data collection burden and improving adaptability to new websites and environments~\cite{chen2022few, chen2021few2, lyu2024tfan, zhou2024joint, wang2024trafficsiam, tan2024adaptability}. Instead of training classifiers from scratch for every target website, few-shot learning models leverage prior knowledge acquired from related tasks or domains to generalize to previously unseen classes with minimal supervision~\cite{xie2024contrastive, li2025cross}.

To improve the efficiency and practicality of DL-based WF attacks, ~\citet{tian2023tiny} proposed TinyWFP, which combines wavelet-based dimensionality reduction with depthwise separable CNNs. By retaining only low-frequency wavelet coefficients, TinyWFP achieved 99.2\% closed-world accuracy while reducing model size and computation by 81× and 79×, respectively, compared with DF. To address limited training data, ~\citet{pan2023practical} combined CNNs with transfer learning and integrated packet direction and timing features, significantly improving performance in few-shot and temporally varying environments. ~\citet{gong2024wfcat} further demonstrated the importance of packet timing by proposing WFCAT, which introduces an inter-arrival time (IAT) histogram representation and a timing-aware CNN architecture. WFCAT achieved over 59\% accuracy against Surakav~\citet{gong2022surakav}, outperforming RF~\citet{shen2023subverting} and Tik-Tok~\citet{rahman2019tik} by 28\% and 48\%, respectively. Despite these advances, existing DL-based WF attacks remain constrained by high computational cost, limited scalability, and high false-positive rates. To address these limitations, ~\citet{chawla2024espresso} proposed ESPRESSO, a Transformer-based flow correlation framework that employs aggregated traffic representations to capture long-range dependencies. Compared with DeepCoFFEA~\cite{oh2022deepcoffea}, ESPRESSO substantially improves correlation accuracy while reducing computational complexity and false positives.

To address concept drift, ~\citet{xu2025apwf} proposed APWF, a parallel WF framework that combines CNNs, BiGRUs, and a convolutional block attention module (CBAM) to jointly capture spatial and temporal traffic features. The authors further incorporated a transfer learning-based fine-tuning strategy to adapt the model to evolving website traffic using only a small amount of labeled data. APWF achieved 92.4\% accuracy when the target data were collected 56 days after the training data, representing only a 4\% performance degradation compared with an immediate attack. ~\citet{ding2024multi} proposed LRCT, a DL-based WF model that combines Local RNN, CNN, and a Transformer Encoder to jointly capture temporal, local spatial, and global traffic features. Experimental results show that LRCT achieves 99.34\% accuracy in the closed-world setting, outperforming prior WF methods while demonstrating improved robustness to concept drift.

Inspired by the success of large language models (LLMs) in sequence modeling, recent studies have begun applying LLMs to website fingerprinting~\cite{zhao2026robust}. These approaches represent traffic traces as tokenized sequences and utilize Transformer-based architectures to learn high-level temporal and structural patterns directly from encrypted traffic. The emergence of LLM-based WF attacks suggests a promising direction for improving feature representation, model transferability, and robustness in realistic deployment scenarios.

\subsubsection{Multi-tab WF attacks}

Early WF attacks are typically developed and evaluated under the single-tab assumption, where each traffic trace corresponds exclusively to a single webpage visit. Under this setting, attackers analyze clean and well-labeled traffic traces without interference from concurrent browsing activities. However, this assumption is often unrealistic in practice, as users frequently engage in multiple online activities simultaneously, such as instant messaging while browsing the web or opening multiple webpages in parallel through different browser tabs~\cite{wang2016realistically, cui2019revisiting, gu2015novel, juarez2014critical}. Early studies demonstrated that such multi-tab browsing behavior substantially degrades the effectiveness of WF attacks designed for single-tab scenarios~\cite{juarez2014critical, wang2016realistically}. Consequently, initial multi-tab WF research focused on decomposing mixed traffic traces into individual webpage segments by identifying traffic split points and subsequently applying conventional single-tab classifiers. Representative approaches employed traffic segmentation~\cite{wang2016realistically, yuan2025mw3f, yin2021automated}, chunk-based classification~\cite{xu2018multi}, and probabilistic models such as Hidden Markov Models~\cite{cui2019revisiting} to recover individual browsing sessions from overlapping traffic traces. Although these methods improved recognition performance, they often suffered from information loss during segmentation, limited scalability as the number of tabs increased, and difficulties in handling highly overlapping traffic patterns~\cite{guan2021bapm}.

Recent advances have shifted from explicit traffic segmentation toward end-to-end deep learning frameworks that directly model overlapping traffic regions. Convolutional neural networks~\cite{cui2020more, cui2023edf, yin2024traces}, attention-based architectures~\cite{xie2023multi}, Transformers~\cite{jin2023transformer, deng2023robust}, state-space models~\cite{liu2024fingermamba}, and metric-learning approaches~\cite{zhao2024towards} have been proposed to capture complex dependencies among concurrent webpage visits while reducing reliance on handcrafted features and prior knowledge of the number of open tabs. More recently, state-space models~\cite{deng2025countmamba} and metric-learning techniques~\cite{zhang2026label} have further improved robustness, early-stage identification capability, and feature discriminability in complex browsing environments. These advances demonstrate a clear transition from segmentation-based solutions to representation-learning approaches, highlighting the growing importance of modeling realistic browsing behaviors and overlapping traffic patterns in modern WF research.

\subsection{WF Defenses}
\label{subsec:bacwfd}

\begin{table}[t]
\centering
\scriptsize
\setlength{\tabcolsep}{3pt}
\renewcommand{\arraystretch}{1.2}
\caption{Taxonomy and characteristics of representative website fingerprinting defense mechanisms.}
\label{tab:wfd}

\begin{tabular}{cccccccc} 
\toprule[0.75pt]
\rowcolor{olive!25}
\multicolumn{2}{c}{\textbf{Categories}}  & \textbf{Defense Model} & \textbf{\makecell{Trace \\ Representation}} & \textbf{\makecell{Adaptive to \\ Live Traffic}} & \textbf{\makecell{Prior knowledge \\ of Traffic}} & \textbf{\makecell{Limited to  \\ Dataset}} & \textbf{Zero Delay}\\ 
\midrule
\rowcolor{cyan!15}
\cellcolor{orange!15}& & WTF-PAD~\cite{juarez2016toward} & Packet Sequence &  \textcolor{green}{\cmark} Yes & \textcolor{red}{\xmark} No & \textcolor{red}{\xmark} No & \textcolor{green}{\cmark} Yes \\
\rowcolor{cyan!15}
\cellcolor{orange!15}& & DynaFlow~\cite{lu2018dynaflow} & Burst Sequence &  \textcolor{green}{\cmark} Yes & \textcolor{red}{\xmark} No & \textcolor{red}{\xmark} No & \textcolor{red}{\xmark} No \\
\rowcolor{cyan!15}
\cellcolor{orange!15}& \multirow{-3}{*}{\makecell{Adaptive \\ Padding}} & FRONT~\cite{gong2020zero} & Packet Sequence & \textcolor{green}{\cmark} Yes & \textcolor{red}{\xmark} No & \textcolor{red}{\xmark} No & \textcolor{green}{\cmark} Yes \\

\rowcolor{gray!15}
\cellcolor{orange!15} & & WF-GAN~\cite{hou2020wf} & Packet Sequence & \textcolor{red}{\xmark} No & \textcolor{green}{\cmark} Yes & \textcolor{green}{\cmark} Yes & \textcolor{red}{\xmark} No \\
\rowcolor{gray!15}
\cellcolor{orange!15}& & DFD~\cite{abusnaina2020dfd} & Burst Sequence & \textcolor{green}{\cmark} Yes & \textcolor{red}{\xmark} No & \textcolor{red}{\xmark} No & \textcolor{red}{\xmark} No \\
\rowcolor{gray!15}
\cellcolor{orange!15}& & Mockingbird~\cite{rahman2020mockingbird} & Packet Sequence & \textcolor{red}{\xmark} No & \textcolor{green}{\cmark} Yes & \textcolor{green}{\cmark} Yes & \textcolor{red}{\xmark} No  \\
\rowcolor{gray!15}
\cellcolor{orange!15}& & BLANKET~\cite{nasr2021defeating}& Packet Sequence & \textcolor{green}{\cmark} Yes & \textcolor{red}{\xmark} No & \textcolor{red}{\xmark} No & \textcolor{red}{\xmark} No \\
\rowcolor{gray!15}
\cellcolor{orange!15}& & Minipatch~\cite{li2022minipatch} & Packet Sequence & \textcolor{green}{\cmark} Yes & \textcolor{green}{\cmark} Yes & \textcolor{red}{\xmark} No & \textcolor{red}{\xmark} No \\
\rowcolor{gray!15}
\cellcolor{orange!15}& & snWF~\cite{wang2022snwf} & Packet Sequence &  \textcolor{green}{\cmark} Yes & \textcolor{green}{\cmark} Yes & \textcolor{red}{\xmark} No & \textcolor{red}{\xmark} No \\
\rowcolor{gray!15}
\cellcolor{orange!15}& & Surakav~\cite{gong2022surakav} & Burst Sequence & \textcolor{green}{\cmark} Yes &  \textcolor{red}{\xmark} No & \textcolor{red}{\xmark} No & \textcolor{red}{\xmark} No \\
\rowcolor{gray!15}
\cellcolor{orange!15}& & ACUP3~\cite{qiao2023resisting} & Burst Sequence & \textcolor{green}{\cmark} Yes & \textcolor{red}{\xmark} No & \textcolor{red}{\xmark} No & \textcolor{red}{\xmark} No \\
\rowcolor{gray!15}
\cellcolor{orange!15}& & ALERT~\cite{qiao2024trace} & Burst Sequence & \textcolor{green}{\cmark} Yes & \textcolor{red}{\xmark} No & \textcolor{green}{\cmark} Yes & \textcolor{red}{\xmark} No \\
\rowcolor{gray!15}
\cellcolor{orange!15}& & RUDOLF~\cite{jiang2024rudolf}& Burst Sequence & \textcolor{green}{\cmark} Yes & \textcolor{red}{\xmark} No & \textcolor{red}{\xmark} No & \textcolor{red}{\xmark} No \\
\rowcolor{gray!15}
\cellcolor{orange!15}& & GAPDiS~\cite{xie2025gapdis} & Packet Sequence & \textcolor{green}{\cmark} Yes & \textcolor{green}{\cmark} Yes & \textcolor{red}{\xmark} No & \textcolor{red}{\xmark} No  \\
\multirow{-15}{*}{\rotatebox[origin=c]{90}{Traffic Obfuscation}} \cellcolor{orange!15}  & \cellcolor{gray!15} \multirow{-12}{*}{\makecell{Adversarial \\ Perturbation}} & \cellcolor{gray!15}  FRUGAL~\cite{wang2026cease} & \cellcolor{gray!15}  Packet Sequence  & \cellcolor{gray!15} \textcolor{green}{\cmark} Yes  & \cellcolor{gray!15} \textcolor{green}{\cmark} Yes & \cellcolor{gray!15} \textcolor{red}{\xmark} No & \cellcolor{gray!15} \textcolor{red}{\xmark} No \\

\rowcolor{cyan!15}
\cellcolor{pink!15} & & BuFLO~\cite{dyer2012peek}  & Packet Sequence & \textcolor{red}{\xmark} No  & \textcolor{green}{\cmark} Yes & \textcolor{red}{\xmark} No &  \textcolor{red}{\xmark} No \\
\rowcolor{cyan!15}
\cellcolor{pink!15}& & CS-BuFLO~\cite{cai2014cs} & Packet Sequence & \textcolor{red}{\xmark} No  & \textcolor{green}{\cmark} Yes & \textcolor{red}{\xmark} No & \textcolor{red}{\xmark} No \\
\rowcolor{cyan!15}
\cellcolor{pink!15}& & Tamaraw~\cite{cai2014systematic} & Packet Sequence & \textcolor{red}{\xmark} No & \textcolor{green}{\cmark} Yes & \textcolor{red}{\xmark} No & \textcolor{red}{\xmark} No \\
\rowcolor{cyan!15}
\cellcolor{pink!15}& & RegularTor~\cite{holland2020regulator} & Packet Sequence &  \textcolor{green}{\cmark} Yes& \textcolor{red}{\xmark} No & \textcolor{red}{\xmark} No & \textcolor{red}{\xmark} No \\
\rowcolor{cyan!15}
\cellcolor{pink!15}& & GLUE~\cite{gong2020zero} & Packet Sequence & \textcolor{green}{\cmark} Yes & \textcolor{red}{\xmark} No & \textcolor{red}{\xmark} No &  \textcolor{green}{\cmark} Yes \\
\rowcolor{cyan!15}
\cellcolor{pink!15}& & Palette~\cite{shen2024real} & Traffic Matrix & \textcolor{green}{\cmark} Yes & \textcolor{red}{\xmark} No & \textcolor{green}{\cmark} Yes & \textcolor{red}{\xmark} No  \\
\rowcolor{cyan!15}
\cellcolor{pink!15}& \multirow{-7}{*}{\makecell{Traffic \\ Regularization}} & Adaptive Tamaraw~\cite{khajavi2026lightening} & Packet Sequence & \textcolor{green}{\cmark} Yes & \textcolor{red}{\xmark} No & \textcolor{red}{\xmark} No & \textcolor{red}{\xmark} No \\

\rowcolor{gray!15}
\cellcolor{pink!15} &  & Supersequence~\cite{wang2014effective} & Packet Sequence & \textcolor{red}{\xmark} No & \textcolor{green}{\cmark} Yes & \textcolor{green}{\cmark} Yes & \textcolor{red}{\xmark} No \\
\rowcolor{gray!15}
\cellcolor{pink!15}& & Walkie-Talkie~\cite{wang2017walkie} & Burst Sequence & \textcolor{green}{\cmark} Yes & \textcolor{green}{\cmark} Yes & \textcolor{green}{\cmark} Yes & \textcolor{red}{\xmark} No \\
\rowcolor{gray!15}
\cellcolor{pink!15}& & BiMorphing~\cite{al2019bimorphing} & Burst Sequence & \textcolor{green}{\cmark} Yes & \textcolor{red}{\xmark} No & \textcolor{green}{\cmark} Yes & \textcolor{red}{\xmark} No \\
\cellcolor{pink!15} \multirow{-11}{*}{\rotatebox[origin=c]{90}{Traffic Regularization}} & \cellcolor{gray!15} \multirow{-4}{*}{\makecell{Traffic \\Morphing}} & \cellcolor{gray!15} WFGuard~\cite{ling2024wfguard} & \cellcolor{gray!15} Packet Sequence & \cellcolor{gray!15} \textcolor{red}{\xmark} No & \cellcolor{gray!15} \textcolor{green}{\cmark} Yes & \cellcolor{gray!15} \textcolor{green}{\cmark} Yes & \cellcolor{gray!15} \textcolor{red}{\xmark} No \\

\bottomrule[0.75pt]
\end{tabular}
\vspace{-0.20in}
\end{table}

WF defenses aim to mitigate information leakage from encrypted traffic by obscuring the characteristics of the traffic that attackers exploit to identify websites. Early WF defense research focused on encrypted communication systems and single-proxy architectures, introducing techniques such as link padding~\cite{fu2003analytical} and traffic regularization~\cite{nithyanand2014glove}. The first defense specifically designed for Tor was proposed by ~\citet{panchenko2011website}, who introduced a camouflage-based approach that generated additional webpage requests to obscure traffic characteristics and confuse attack classifiers. Subsequent research increasingly shifted toward defending against Tor-based WF attacks, driven by the growing effectiveness of machine-learning and deep-learning classifiers.

Existing WF defenses can be broadly categorized into two fundamental classes: traffic obfuscation and traffic regularization. Traffic obfuscation seeks to conceal website-specific characteristics by introducing perturbations into network traffic, such as dummy packet injection and traffic burst modifications~\cite{juarez2016toward, gong2020zero, hou2020wf, rahman2020mockingbird, nasr2021defeating, wang2026cease}, thereby reducing the amount of exploitable information available to an attacker. In contrast, traffic regularization aims to transform traffic traces from different websites into more homogeneous patterns, minimizing inter-class variability and making websites harder to distinguish~\cite{holland2020regulator, wang2017walkie, khajavi2026lightening}. Building upon these two underlying design principles, this survey further classifies existing WF defenses into four representative categories according to their operational mechanisms and design objectives: \textbf{adaptive padding}, \textbf{adversarial perturbation}, \textbf{traffic regularization}, and \textbf{traffic morphing}-based defenses. This taxonomy provides a unified perspective for analyzing the effectiveness, deployment overhead, and robustness of defense mechanisms against modern WF attacks.

\subsubsection{Adaptive Padding}

Adaptive traffic padding is a technique used to protect against WF attacks by dynamically or randomly injecting dummy packets and introducing controlled delays into network traffic streams~\cite{shmatikov2006timing}. The core idea of this defense mechanism is to obfuscate traffic characteristics that are commonly exploited by WF classifiers, such as burst patterns~\cite{lu2018dynaflow} and packet ordering~\cite{juarez2016toward, gong2020zero}. Unlike static padding schemes~\cite{panchenko2011website, cai2012touching}, adaptive padding selectively modifies traffic according to observed communication patterns, enabling a more favorable trade-off between privacy protection and performance overhead. Existing approaches generally employ traffic obfuscation techniques, such as inserting dummy packets into statistically unlikely timing gaps, generating realistic cover traffic, or perturbing burst structures through randomized padding strategies. Representative defenses, including WTF-PAD~\cite{juarez2016toward}, FRONT and GLUE~\cite{gong2020zero}, DFD~\cite{abusnaina2020dfd}, and RBP~\cite{luo2021rbp}, demonstrate that adaptive padding can substantially reduce the effectiveness of both traditional machine learning and deep learning-based WF attacks while incurring moderate bandwidth and latency overhead. Subsequent studies further explored user-specific cover traffic generation, probabilistic padding mechanisms, and circuit-level padding strategies to improve defense effectiveness under realistic deployment conditions.

Recent research has focused on improving the practicality, adaptability, and robustness of adaptive padding defenses. Techniques such as RanDePad~\cite{hong2022website}, TBP~\cite{wang2023tbp}, WFD$^{3}$P~\cite{lu2023lightweight}, AAP~\cite{yang2023aap}, BRO~\cite{mcguan2024practical}, Maybenot~\cite{pulls2023maybenot}, and Break-Pad~\cite{huang2024break} employ increasingly sophisticated traffic perturbation strategies, including adaptive random delays, burst-sequence padding, bursts distribution-aware packet injection, probabilistic packet generation, and protocol-agnostic defense frameworks. These methods seek to maximize uncertainty for attackers while maintaining low deployment costs and minimal impact on user experience. Overall, the evolution of adaptive padding defenses reflects a shift from simple dummy-packet insertion toward intelligent, traffic-aware perturbation mechanisms that balance security, bandwidth overhead, latency, and deployability. Despite significant progress, designing adaptive padding schemes that remain effective against increasingly powerful and adaptive WF attacks while preserving network efficiency continues to be an open research challenge.
    
\subsubsection{Traffic Regularization}
 
Although adaptive padding defenses can effectively reduce the success rate of WF attacks by injecting dummy packets and perturbing traffic patterns, recent studies have shown that advanced DL-based WF attacks can still extract discriminative features through sophisticated temporal analysis and representation learning techniques~\cite{wang2021one}. This limitation largely stems from the fact that many padding mechanisms follow deterministic or statistically predictable patterns, allowing powerful classifiers to learn residual traffic characteristics despite the added noise. To address this challenge, traffic regularization defenses adopt a fundamentally different design philosophy. Rather than concealing traffic features through packet manipulation alone, these approaches seek to regulate traffic representations such that traces generated by different websites exhibit similar burst characteristics. By reducing inter-class variability, regularization-based defenses hinder the ability of machine learning models to learn distinctive website fingerprints, thereby mitigating both traditional and DL-based WF attacks.

Early regularization-based defenses focused on standardizing traffic transmission patterns through fixed-rate communication. ~\citet{dyer2012peek} introduced Buffered Fixed-Length Obfuscation (BuFLO), which transmits fixed-size packets at fixed intervals to conceal traffic volume and timing characteristics. Although BuFLO significantly reduced the effectiveness of several WF attacks, its substantial bandwidth and latency overhead limited practical deployment. Subsequent improvements, including Congestion-Sensitive BuFLO (CS-BuFLO)~\cite{cai2014cs} and Tamaraw~\cite{cai2014systematic}, sought to improve the trade-off between security and communication overhead through adaptive transmission rates and more efficient traffic shaping mechanisms. More recent defenses have shifted toward learning-based traffic regularization strategies. RegulaTor~\cite{holland2020regulator} leverages common traffic patterns observed in web browsing sessions to regulate traffic traces while maintaining practical overhead. Similarly, BM~\cite{chen2023web} incorporates temporal information into burst shaping by constructing unified burst sequences that obscure website-specific traffic patterns. Palette~\cite{shen2024real} further advances this direction by introducing traffic cluster anonymization, which groups websites with similar traffic characteristics and regulates their traffic traces toward a common profile. These approaches demonstrate a clear evolution from fixed-rate traffic shaping toward adaptive and representation-aware regularization mechanisms that aim to provide stronger protection against modern DL-based WF attacks while maintaining practical deployment costs. However, Palette is highly dependent on the availability of representative training datasets and the stability of website traffic distributions over time. To address this limitation, \citet{khajavi2026lightening} proposed a unified framework for adaptive WF defense that combines the traffic homogenization benefits of regularization with the provable security guarantees of supersequence-based grouping.

\subsubsection{Traffic Morphing}

Traffic morphing is a traffic transformation technique designed to mitigate WF attacks by altering the observable characteristics of network traffic such that the burst traffic generated by one website resembles that of another~\cite{wright2009traffic, nithyanand2014glove}. Unlike padding-based defenses that primarily inject dummy packets or delays, traffic morphing directly modifies traffic representations to mislead classifiers. The underlying objective is to reduce the distinguishability of website-specific traffic patterns by transforming traffic traces toward a target profile or a shared representation. Early work in this direction includes Supersequence~\cite{wang2014effective}, which transforms webpage traces into a common supersequence shared among a set of websites, thereby preventing attackers from distinguishing individual webpages within the anonymity set. Subsequently, Walkie-Talkie~\cite{wang2017walkie} introduced burst molding and half-duplex communication to make sensitive and non-sensitive webpages appear indistinguishable. By reshaping traffic bursts rather than relying solely on packet padding, Walkie-Talkie achieved a more favorable trade-off between security and overhead. However, it requires prior knowledge of target webpages and is less effective for highly dynamic content.

Subsequent research further improved traffic morphing through clustering, optimization, and learning-based techniques. \citet{chan2018website} proposed clustering websites according to traffic characteristics and morphing all traces within a cluster to resemble a representative target, thereby reducing intra-cluster distinguishability. BiMorphing~\cite{al2019bimorphing} extended this concept by leveraging bidirectional burst sampling and optimization strategies to achieve strong protection with low bandwidth overhead and no additional latency. More recently, researchers have explored traffic morphing defenses specifically designed to counter DL-based WF attacks. For example, WFGuard~\cite{ling2024wfguard} formulates traffic transformation as an adversarial optimization problem, utilizing fine-grained neuron feedback from WF classifiers to iteratively generate traffic mutations that maximize misclassification. By combining gradient-based traffic modification with dummy packet injection, WFGuard substantially reduces the effectiveness of prior proposed DL attacks while maintaining relatively low bandwidth overhead. Overall, the evolution of website morphing defenses reflects a transition from rule-based traffic transformation to adaptive, optimization-driven approaches that explicitly target the feature-extraction mechanisms employed by modern WF classifiers.

\subsubsection{Adversarial Perturbation}

Adversarial perturbation-based defenses have emerged as a promising direction for mitigating WF attacks, particularly against modern DL-based classifiers. Unlike traffic padding, regularization, or morphing approaches that modify traffic according to predefined rules, adversarial defenses leverage techniques from adversarial machine learning to generate carefully crafted perturbations that induce misclassification by WF models~\cite{moosavi2017universal, huang2025wf}. The fundamental objective is to manipulate traffic representations while preserving communication functionality, thereby reducing the ability of machine learning models to extract discriminative website fingerprints. Early studies demonstrated the feasibility of this approach by generating adversarial traffic traces through dummy packet injection and traffic mutation strategies that intentionally shifted classifier predictions toward incorrect website classes~\cite{imani2018adversarial, li2019dynamic, cai2020jumpestimate}. Subsequent work extended these ideas through generative adversarial networks (GANs), boundary-attack-based optimization, and trace transformation techniques capable of deceiving both traditional machine learning and deep learning-based WF classifiers. Representative defenses such as WF-GAN~\cite{hou2020wf}, Mockingbird~\cite{rahman2020mockingbird}, PST~\cite{jiang2020pst}, Dolos~\cite{shan2021patch}, and universal adversarial perturbation methods~\cite{hou2021universal, sun2022practical} demonstrated that adversarial perturbations can substantially reduce attack accuracy while incurring relatively low bandwidth overhead.

Recent research has focused on improving the practicality, transferability, and robustness of adversarial defenses under realistic deployment constraints. Many early approaches relied on strong assumptions, such as access to complete traffic traces, white-box knowledge of the attack model, or vulnerability to adversarial retraining. To overcome these limitations, later studies proposed lightweight and black-box defenses that generate perturbations in real time while maintaining low communication overhead. Examples include Minipatch~\cite{li2022minipatch}, TED~\cite{jiang2022effective}, Prism~\cite{li2023prism}, Acup3~\cite{qiao2023resisting}, KimeraPad~\cite{jiang2024kimerapad}, CMAES-WFD~\cite{wang2024cmaes}, ALERT~\cite{qiao2024trace}, and RUDOLF~\cite{jiang2024rudolf}. These methods employ diverse optimization strategies, including evolutionary algorithms, reinforcement learning, universal perturbation generation, and traffic-aware adversarial adaptation, to achieve effective protection without requiring extensive prior knowledge of traffic traces or attack models. More recently, approaches such as GAPDiS~\cite{xie2025gapdis} have combined gradient-guided optimization with global search mechanisms to generate highly transferable perturbations that remain effective across a wide range of websites and attack settings. Overall, the evolution of adversarial perturbation-based defenses reflects a transition from trace-specific and model-dependent perturbations toward practical, adaptive, and transferable defense mechanisms that can effectively counter increasingly sophisticated DL-based WF attacks while maintaining low deployment overhead.

\subsection{Datasets}
\label{subsec:dataset}

\begin{table}
\centering
\scriptsize
\setlength{\tabcolsep}{3pt}
\renewcommand{\arraystretch}{2.2}
\caption{Characteristics of representative public datasets for website fingerprinting research. More dataset comparisons and download links are available at \url{https://github.com/WF-Attack-and-Defense/Awesome-WF-ecurity/blob/main/datasets.md}.}
\label{tab:publicdataset}
\begin{tabular}{cccccccc} 
\toprule[0.75pt]
\textbf{Dataset Name} & \textbf{Year}  & \textbf{User Model} & \textbf{\makecell{Traffic \\ Characteristics}} & \textbf{\makecell{Included \\ Multi-tab}} & \textbf{\makecell{Dataset \\ Type}} & \textbf{\makecell{Dataset Size \\ $N_{C} \times N_{I}$}} \\ 
 \midrule
AWF~\cite{rimmer2017automated} & 2017 & Index page &  \makecell{Traffic direction} & \xmark & Synthetic & \parbox[c]{2.3cm}{\centering CW: 900$\times$2500 \\ OW: 400,000+200} \\
\rowcolor{gray!25}
Walkie-Talkie~\cite{wang2017walkie} & 2017 & Index page & Traffic timing, Traffic direction  & \xmark & Synthetic  &  \parbox[c]{2.3cm}{\centering CW: 100$\times$100  \\ OW: 9,900$\times$1} \\
DF~\cite{sirinam2018deep} & 2018 & 	Index page & \makecell{ Traffic timing,  Packet Size, \\ Traffic direction}  & \xmark & Synthetic & \parbox[c]{2.3cm}{\centering CW: 95$\times$1000 \\ OW: 40,716$\times$1} \\
\rowcolor{gray!25}
DS-19~\cite{gong2020zero} & 2020 & Index page & Traffic timing, Traffic direction & \xmark  & Synthetic & \parbox[c]{2.3cm}{\centering CW: 100$\times$100 \\ OW: 10,000$\times$1} \\
ARES~\cite{deng2023robust} & 2023 &  Index page & Traffic timing, Traffic direction & \cmark & Synthetic & \parbox[c]{2.3cm}{\centering CW: 100$\times$1,000 \\ OW: 100$\times$1,000} \\
\rowcolor{gray!25}
TMWF~\cite{jin2023transformer} & 2023 & Index page & Traffic timing, Traffic direction & \cmark & Synthetic  & \parbox[c]{2.3cm}{\centering CW: $\text{TBB}_{S}$: 50$\times$100 \\ OW: $\text{TBB}_{S}$: 5,000$\times$1 \\ OW: $\text{TBB}_{M}$: 10,000$\times$1} \\
GTT23 ~\cite{jansen2026measurement} & 2023 & Visited service & Traffic timing, Traffic direction & \xmark & Genuine  & \parbox[c]{2.3cm}{\centering $1.4 \times 10^{7}$ \\ ($1.1 \times 10^{6}$ domains)} \\
\rowcolor{gray!25}
Oscar ~\cite{zhao2024towards} & 2024 & Index \& Subpage & Traffic timing, Traffic direction & \cmark & Synthetic  & \parbox[c]{2.3cm}{\centering CW: 1,000 $\times$ 10 \\ OW: 9,236 $\times$1} \\
\bottomrule[0.75pt]
\multicolumn{6}{l}{\makecell[l]{CW: Closed World dataset; OW: Open World dataset; $S$: Single-tab dataset; $M$: Multi-tab dataset. \\ $N_{C}$: number of dataset classes; $N_{I}$: number of instances per class.}} \\
\end{tabular}
\vspace{-0.20in}
\end{table}

The evaluation of WF attacks and defenses is fundamentally dependent on the datasets used for model development and assessment. Existing WF datasets can be broadly categorized into \emph{synthetic} and \emph{genuine} datasets according to their collection methodology. Synthetic datasets are generated in controlled environments with carefully orchestrated webpage visits, whereas genuine datasets capture real-world browsing behavior and its associated variability. Synthetic datasets can be further divided into closed-world and Open-world settings~\cite{wang2014effective, panchenko2016website, rimmer2017automated}, which differ in their threat assumptions, dataset composition, and evaluation objectives. Together, these dataset categories provide complementary perspectives for evaluating the effectiveness, robustness, and practical applicability of WF attacks and defenses.

\subsubsection{Synthetic Datasets}

Synthetic datasets are collected in controlled experimental environments using automated traffic collection frameworks, such as the Tor Browser Bundle (TBB)~\cite{torbrowser35}, Tor-Browser-Crawler~\cite{juarez2014critical}, and Tor-Browser-Selenium~\cite{tor-browser-selenium}. These frameworks automate webpage visits to generate labeled traffic traces under predefined conditions, enabling reproducible and large-scale dataset collection. Since the traffic is typically generated by a single crawling application following fixed browsing behaviors, synthetic datasets generally capture only simple webpage access patterns and lack the diversity and unpredictability of real user browsing activities~\cite{jansen2026measurement}. Most existing synthetic datasets are collected by monitoring the encrypted traffic between the Tor client and the entry (guard) relay, which reflects the adversarial setting commonly assumed in WF research.

\emph{Closed-World Datasets:} The closed-world setting assumes that users visit only a predefined set of websites, referred to as \emph{monitored websites}~\cite{cai2012touching, wang2016realistically, hayes2016k}. Under this assumption, the attacker has complete knowledge of all candidate websites and seeks to classify each traffic trace into one of the monitored classes. Consequently, the task is formulated as a multi-class classification problem over a finite set of websites. Although this setting does not fully reflect real-world browsing behavior, it provides a controlled and reproducible evaluation environment for assessing the discriminative power of traffic features and the effectiveness of WF attack models. As a result, closed-world datasets are widely adopted for benchmarking newly proposed attacks and comparing different model architectures under consistent experimental conditions~\cite{sirinam2019triplet}. Performance achieved in the closed-world setting is generally regarded as an upper bound on the effectiveness of WF attacks.

\textit{Open-World Datasets:} The open-world setting provides a more realistic threat model by considering a relatively small set of monitored websites together with a substantially larger collection of unmonitored websites~\cite{wang2014effective}. Rather than identifying every visited website, the attacker's objective is to determine whether a traffic trace belongs to a monitored website or should be classified as unmonitored~\cite{wang2013improved}. This setting more closely reflects practical browsing behavior, where users may access an effectively unlimited number of websites. A fundamental challenge in open-world evaluation is the base-rate fallacy, in which even attacks with high detection rates can generate many false positives because unmonitored traffic vastly outnumbers monitored traffic. Consequently, open-world evaluations typically report metrics such as the TPR, FPR, precision, recall, and F1-score to provide a comprehensive assessment of attack performance.

\subsubsection{Genuine Datasets}

Although synthetic datasets have been widely adopted for developing and evaluating WF attacks and defenses, they cannot fully capture the diversity and complexity of real-world Tor browsing behavior. As discussed by \citet{jansen2026measurement}, traffic collected in controlled environments often differs from genuine user activities in terms of browsing patterns, website popularity, background traffic, and network dynamics, which may introduce biases into experimental evaluations. To address this limitation, the authors presented the first large-scale genuine WF dataset collected from a Tor exit relay, where traffic naturally reflects real Tor users accessing websites over the public Internet. The resulting dataset contains approximately (1.4$\times10^{7}$) website visits spanning 91 days and covering more than (1.1$\times10^{6}$) unique website domains. Unlike synthetic datasets, which rely on automated crawlers visiting a predefined set of websites, this dataset captures realistic browsing behaviors without artificial traffic generation. Furthermore, it can be partitioned into both closed-world and open-world settings to support the evaluation of diverse WF attack and defense models. To the best of our knowledge, this is currently the only publicly reported genuine dataset specifically designed for WF research, providing an important benchmark for assessing the robustness and practical applicability of WF techniques under realistic browsing conditions.

\tableautorefname~\ref{tab:publicdataset} summarizes representative public datasets that have been widely adopted for the development and evaluation of website fingerprinting (WF) attacks and defenses. The comparison illustrates a clear evolution in dataset design over the past decade, progressing from relatively small, controlled datasets toward larger, more diverse, and increasingly realistic benchmarks. Early datasets primarily targeted single-tab browsing under the closed-world assumption, providing reproducible environments for evaluating the discriminative capability of proposed attack models. More recent datasets have expanded to include large-scale open-world settings, multi-tab browsing scenarios, and richer traffic characteristics, reflecting the growing emphasis on evaluating WF techniques under practical browsing conditions. Across these datasets, traffic timing and traffic direction remain the most commonly used observable features, highlighting their central role in modern WF attacks. Although genuine datasets remain scarce because of the challenges associated with large-scale collection, privacy preservation, and data annotation, they represent an important direction for future WF research.

\section{Open Challenges and Future Research Directions}
\label{sec:challenges}

In this section, we discuss the major open challenges and future research directions in WF research by synthesizing insights from the surveyed literature. We examine unresolved issues spanning datasets, attack models, defense mechanisms, and evaluation methodologies, highlighting the limitations that hinder fair comparison, practical deployment, and long-term robustness. Based on these observations, we identify promising research directions that can advance the development of more realistic, effective, and deployable WF techniques for Tor and other anonymous communication systems.

\subsection{Challenges in WF Attacks}

Although WF attacks have achieved remarkable progress in recent years, their practical deployment remains challenging. Most existing approaches are developed and evaluated under controlled experimental settings that rely on strong assumptions regarding user behavior, network conditions, and attacker capabilities. In real-world environments, factors such as evolving website content, network variability, realistic browsing behaviors, and limited training data can substantially degrade attack performance. This subsection discusses the major challenges facing current WF attacks, including effectiveness, robustness, generalizability, and transferability, and highlights the key limitations that must be addressed to enable practical and reliable deployment.

\subsubsection{Effectiveness.} The primary goal of WF attacks is to correctly distinguish traffic traces belonging to different websites. This requires representations that simultaneously maximize intra-class similarity and inter-class separability. Existing methods typically rely on low-level traffic characteristics, including packet direction, timing, ordering, burst patterns, and frequency. Although these approaches achieve strong performance in controlled settings, most representations remain limited to low-level statistical features of encrypted traffic. A promising direction is the exploration of higher-level behavioral or semantic representations that better capture interactions among clients, relays, and servers. Such representations may improve discriminability and robustness across diverse network conditions. Additionally, many WF attacks are evaluated in closed-world settings~\cite{shen2019webpage}, where all test traces belong to a small, fixed set of monitored websites. However, real-world scenarios are inherently open-world, where monitored sites constitute only a small fraction of all traffic. Effectively distinguishing monitored traffic from a large volume of unmonitored traffic remains challenging, particularly under modern defense mechanisms that deliberately obfuscate discriminative patterns.


\subsubsection{Robustness.} Robustness refers to the ability of WF attacks to maintain performance under distribution shifts in traffic data. These shifts may arise from website updates, protocol evolution, network variability, user behavior changes, or deployed defenses. Many handcrafted features are highly sensitive to such variations, leading to significant performance degradation under mismatched training and testing conditions. In contrast, sequence-based and representation learning approaches, such as those based on packet streams, graph structures, or traffic images, tend to be more resilient due to their ability to preserve richer structural information. However, designing representations that remain stable under diverse and evolving network conditions remains an open challenge. Future work should focus on learning invariant representations that capture fundamental traffic properties while reducing sensitivity to environmental noise.

\subsubsection{Generalizability.} Generalizability refers to a model's ability to maintain high performance on unseen data without requiring retraining. Existing WF studies often assume that training and testing data are collected under similar conditions, including comparable network characteristics, browser configurations, operating systems, and user behaviors. Such assumptions can lead to overly optimistic performance estimates. In practice, attackers operate in heterogeneous environments where traffic distributions vary substantially across networks, devices, browsers, and time. Consequently, a model trained under one set of conditions may experience significant performance degradation when directly applied to another environment. Improving cross-environment generalization is therefore essential for developing WF attacks that remain effective under realistic deployment scenarios.

\subsubsection{Transferability.} Transferability refers to a model's ability to leverage previously learned knowledge to efficiently adapt to new datasets, websites, tasks, or deployment environments with minimal additional training. Unlike generalization, which evaluates performance without modifying the model, transferability focuses on reducing the cost of adapting an existing model to new conditions. Most supervised WF attacks learn dataset-specific representations and fixed mappings between traffic traces and website labels. As websites evolve or new environments emerge, these models typically require costly retraining and new labeled datasets. Promising directions include transfer learning, domain adaptation, continual learning, self-supervised learning, and foundation-model-based approaches, which aim to reuse learned representations and minimize adaptation effort while maintaining high classification performance.

\subsubsection{Dataset Limitations} The effectiveness of WF attacks depends heavily on the availability of representative training datasets. Modern DL-based WF attacks typically require massive amounts of labeled traffic traces to achieve high classification accuracy. Consequently, attackers must continuously collect and label large-scale datasets that reflect evolving websites, network conditions, and user browsing behaviors, making dataset construction both time-consuming and resource-intensive. Although few-shot and low-shot learning techniques have been proposed to reduce the amount of required training data, their performance remains significantly below that of fully supervised approaches and is not yet sufficient for realistic large-scale deployment. Furthermore, existing public datasets are often limited in scale, diversity, browsing realism, and collection duration, causing models trained on them to overestimate real-world performance. In particular, realistic multi-tab browsing, long-term website evolution, and representative user behavior remain underrepresented in current datasets~\cite{shen2022machine}.

\subsubsection{Model Drift} Model drift denotes the degradation of a machine learning model over time as the deployment distribution diverges from the training distribution, due to shifts in data statistics, input–output relationships, or environmental conditions. In the WF literature, model drift is typically conflated with two distinct phenomena: \emph{data drift} and \emph{concept drift}. Although often used interchangeably, these capture fundamentally different sources of performance degradation and require different mitigation strategies.

In WF settings, this arises when website traffic patterns evolve (\eg, dynamic content, personalization, third-party resources) without altering website identity. Although packet-level features such as burst patterns, timing, and volume may shift, label semantics remain stable under standard domain-level dataset construction. Consequently, many effects previously attributed to concept drift in WF studies are more accurately characterized as data drift. Such drift primarily induces distribution mismatch, typically addressable via retraining, augmentation, or domain adaptation.

In WF, true concept drift occurs when the semantic correspondence between traffic patterns and website identities changes (\eg, major architectural redesigns, protocol transitions, or service reconfiguration) that fundamentally alters traffic generation mechanisms. Unlike data drift, concept drift implies instability of the labeling function itself, requiring continual or adaptive learning beyond static retraining~\cite{wang2022snwf}.

\subsection{Challenges of WF Defenses}

The development of WF defenses has largely followed a reactive paradigm, where new defenses are introduced in response to increasingly powerful attacks. This asymmetry creates a persistent challenge for anonymity systems, as defenders must anticipate future attack capabilities while simultaneously preserving network usability and performance. Moreover, many existing defenses are evaluated under constrained experimental settings that do not fully capture real-world deployment complexity. As a result, reported effectiveness may not accurately reflect operational performance. Based on our survey, several fundamental challenges continue to limit the practicality and robustness of WF defenses.

\subsubsection{Balancing Privacy, Bandwidth, and Latency} A fundamental challenge in WF defense design is balancing privacy protection with system overhead~\cite{mcguan2024practical, khajavi2026lightening}. Most defenses reduce fingerprintability by introducing traffic obfuscation~\cite{witwer2022padding, panchenko2011website, cai2012touching, wang2014effective, xie2025gapdis} and traffic regularization~\cite{holland2020regulator, wang2017walkie, shen2024real}. While effective in obfuscating traffic patterns, these techniques inevitably increase bandwidth consumption, latency, and computational cost. The trade-off is particularly acute in Tor, where usability depends on maintaining acceptable browsing performance. Stronger defenses typically require higher perturbation levels, resulting in increased resource consumption and degraded user experience. In contrast, lightweight defenses often fail to provide sufficient protection against SOTA attacks. Designing adaptive mechanisms that dynamically balance privacy guarantees and performance constraints remains an open research problem.

\subsubsection{Lack of Standardized and Realistic Evaluation Frameworks} A major limitation in current WF defense research is the absence of standardized evaluation methodologies. Existing defense studies often adopt different datasets, threat models, attack implementations, network conditions, and evaluation metrics, making direct comparison difficult~\cite{rahman2020mockingbird, jiang2020pst, nasr2021defeating, sun2022practical}. In addition, many evaluations rely on simplified assumptions about user behavior and network conditions, such as perfectly segmented traces, homepage-only browsing, fixed numbers of concurrent tabs, or static webpage content~\cite{ling2024wfguard, gong2022surakav}. These assumptions can substantially overestimate defense effectiveness.

Another limitation is the narrow focus of evaluation metrics. Many studies primarily report attack accuracy while overlooking other critical dimensions, including information leakage, bandwidth and latency overhead, computational cost, energy consumption, and long-term robustness~\cite{li2018measuring}. Developing standardized benchmarks, shared datasets, and unified evaluation frameworks is essential for enabling fair and reproducible comparison across defenses.

\subsubsection{Robustness Against Adaptive Adversaries} Most WF defenses are evaluated against static attack models trained before deployment. In practice, however, adversaries can adapt by collecting new traffic under defended conditions and retraining their models~\cite{khajavi2026lightening}. This adaptive capability can significantly reduce the effectiveness of defenses that appear strong under non-adaptive evaluation settings. As a result, evaluating defenses only against fixed attackers may lead to overly optimistic conclusions. Future evaluations should explicitly consider adaptive adversaries, including adversarial retraining, continual learning, and defense-aware attack strategies. Designing defenses that remain effective under such adaptive scenarios remains a key open challenge.

\subsubsection{Deployment in Real-World Environments} Although many WF defenses achieve strong performance in controlled environments, relatively few are practical for large-scale deployment. Many approaches rely on computationally expensive optimization procedures, learning-based mechanisms, or complex traffic scheduling strategies that are difficult to deploy on resource-constrained devices~\cite{xie2025gapdis, jiang2024kimerapad, qiao2024trace, gong2022surakav}. Beyond computational cost, real-world deployment introduces additional challenges, including protocol compatibility, scalability across heterogeneous devices, maintenance overhead, and user acceptance. Increased latency, bandwidth usage, and energy consumption can further degrade usability and hinder adoption. As a result, a significant gap remains between research prototypes and deployable defense systems. Future work should prioritize lightweight, scalable, and deployable solutions that provide meaningful privacy guarantees under realistic constraints.

\subsubsection{Generalization Across Attack Models} Another key challenge is ensuring that defenses remain effective against diverse and evolving attack models. Many existing defenses are evaluated against a limited set of representative attacks available at the time of publication. However, the rapid evolution of WF techniques means that defenses can quickly become outdated when faced with new architectures, feature representations, or learning paradigms.

Recent advances in deep learning, self-supervised learning, graph neural networks, transformers, and foundation models have significantly expanded the attack landscape~\cite{shi2025multiscale, zou2024relation, zhao2024towards, yuan2025mw3f, gong2024wfcat}. Consequently, defenses designed for traditional classifiers may no longer provide sufficient protection against modern approaches.

Future research should therefore prioritize generalization and robustness rather than optimization against specific attack implementations. Evaluating defenses across diverse attack families and emerging learning paradigms will be essential for ensuring long-term effectiveness.

\subsection{Challenges in WF Datasets}

Traffic datasets form the foundation of WF research, as they are used to train, validate, and benchmark both attack and defense models. Consequently, the quality, diversity, and representativeness of these datasets have a direct impact on the reliability, robustness, and generalizability of experimental results. However, existing WF datasets often fail to capture the complexity of real-world browsing environments, including evolving websites, diverse user behaviors, heterogeneous network conditions, and long-term temporal changes. These limitations can lead to overly optimistic performance estimates and hinder the development of practical WF solutions. This section discusses the major challenges associated with current WF datasets and their implications for future research.

\subsubsection{Challenges in Synthetic Datasets}

Most publicly available WF datasets are synthetically generated via automated web crawling in carefully controlled experimental settings~\cite{jansen2026measurement}. Such datasets typically rely on several simplifying assumptions regarding user browsing behavior, webpage characteristics, and network environments. Common assumptions include the index-page assumption~\cite{herrmann2009website, rimmer2017automated, sirinam2018deep, gong2020zero}, static webpage assumption~\cite{cai2012touching, wang2013improved, wang20202ch, jin2023transformer,  pan2023practical, xu2025apwf}, passive webpage assumption~\cite{panchenko2011website, rimmer2017automated, sirinam2018deep, gong2020zero}, cache-disabled assumption~\cite{gong2020zero, panchenko2016website, hayes2016k, wang2017walkie}, traffic trace replicability, consistent traffic parsing, and multi-tab browsing with fixed overlap ratios~\cite{aminuddin2023rise, jin2023transformer, deng2023robust}. While these assumptions facilitate large-scale dataset collection and reproducible experimentation, they substantially simplify real-world browsing scenarios and may inadvertently overestimate an adversary's capabilities. As a result, models trained and evaluated on synthetic datasets may exhibit degraded performance when deployed in practical environments.

\emph{Index-page assumption.} The index-page assumption is one of the most prevalent simplifications adopted in synthetic WF dataset construction. Under this assumption, each traffic trace corresponds to a homepage visit on a monitored website, and the adversary is assumed to classify each isolated page visit. Although this setting simplifies dataset collection and model evaluation, it does not accurately reflect realistic browsing behavior~\cite{wang2013improved}. In practice, users rarely visit only a website's homepage; instead, they navigate through multiple interconnected pages, generating sequences of traffic traces with diverse content, dynamic resource loading, and varying temporal dependencies. Consequently, homepage-only datasets fail to capture the traffic diversity observed in real browsing sessions, limiting WF models' ability to generalize to practical deployment scenarios.

\emph{Static webpage assumption.} Many existing WF datasets assume that the content and structure of monitored webpages remain relatively stable over time, allowing classifiers trained on historical traffic traces to maintain their effectiveness without frequent retraining. This assumption simplifies long-term dataset collection and evaluation by minimizing temporal variations in webpage traffic. However, it is rarely satisfied in practice. Modern websites, particularly news portals, social media platforms, and video streaming services, continuously update their content, embedded resources, advertisements, and webpage structures~\cite{wang2013improved}. These changes introduce temporal variations in network traffic, resulting in a distribution shift between previously collected training data and newly observed traffic traces. Consequently, classifiers trained on outdated datasets often experience significant performance degradation as webpages evolve, requiring continuous dataset collection and model retraining to maintain high classification accuracy.

\emph{Passive webpage assumption.} Many existing WF datasets assume that webpages are passive, meaning that dynamic or actively generated content is disabled or excluded during data collection. \citet{aminuddin2023rise} define active content as dynamically generated webpage elements, including JavaScript-driven resources, Flash objects, and Java applets. In practice, this assumption is often extended to disabling advertisements, third-party plugins, personalized content, and other context-dependent webpage modifications (\eg, seasonal or location-specific content). Such simplifications reduce traffic variability and improve the reproducibility of collected traces. However, modern webpages are inherently dynamic, with content that frequently changes according to user interactions, browser states, geographic locations, advertising services, and real-time updates. Consequently, repeated visits to the same webpage may generate substantially different traffic patterns even under identical network conditions. Datasets collected under the passive webpage assumption, therefore, fail to capture the variability of realistic webpage traffic, potentially leading to overly optimistic evaluations of WF attacks and reduced model generalizability.

\emph{Cache-disabled assumption.} Many existing WF datasets are collected with browser caching disabled, ensuring that all webpage resources are downloaded upon every visit. This assumption simplifies dataset construction by improving trace reproducibility and reducing variability across repeated measurements. However, it does not accurately reflect realistic browsing behavior. Although the Tor Browser clears certain cached data between browsing sessions to enhance user privacy, cached resources may still be retained and reused within the same session. Consequently, repeated visits to the same webpage can generate substantially different traffic patterns depending on the browser's cache state, the availability of cached resources, and the webpage's resource update behavior.

\emph{Replicability assumption.} Most existing WF datasets assume that an attacker can closely replicate the system and network environment of the target user when collecting training data. This includes reproducing factors such as the browser configuration, operating system, hardware platform, network bandwidth, latency, and other environmental conditions. Such an assumption simplifies dataset construction by minimizing distributional differences between the training and testing data, thereby facilitating the development and evaluation of WF classifiers. However, accurately replicating a target user's environment is rarely feasible in practice. User environments are inherently heterogeneous and continuously evolve due to software updates, hardware diversity, network fluctuations, browser configuration changes, and varying browsing behaviors.

\emph{Traffic parsing assumption.} Most existing WF datasets assume that webpage traffic can be accurately segmented into individual traces before feature extraction and classification. Under this assumption, the attacker is assumed to know the precise start and end of each webpage loading session within a continuous Tor traffic stream. Such an assumption greatly simplifies dataset construction and model evaluation by providing perfectly labeled and isolated traffic traces. However, it is rarely satisfied in real-world deployments. In practice, Tor traffic consists of continuous streams of fixed-size cells in which webpage traffic is interleaved with background activities, persistent connections, browser prefetching, and concurrent webpage requests. These factors obscure session boundaries, making accurate traffic parsing a challenging task.

\emph{Fixed Overlap Ratio in Multi-tab Browsing.} To model multi-tab browsing behavior, many studies construct datasets by artificially overlapping traffic traces according to predetermined ratios, such as 10\%, 20\%, or 50\%~\cite{guan2021bapm, jin2023transformer, deng2023robust, deng2025countmamba}. This strategy provides a simple and reproducible framework for evaluating WF attacks under controlled conditions. However, fixed overlap ratios represent only a coarse approximation of real-world browsing behavior. In practice, users dynamically open, close, refresh, and switch between browser tabs, resulting in highly variable and irregular traffic interactions. The extent of traffic overlap depends on numerous factors, including webpage loading latency, user interaction patterns, browser scheduling, background processes, and network conditions. Consequently, the temporal relationships among concurrent traffic flows are highly dynamic and cannot be accurately characterized by a small set of predefined overlap ratios.

\emph{User Browsing Behavior Assumptions} In addition to assumptions regarding webpage characteristics and traffic generation, many WF datasets are constructed under simplified assumptions about user browsing behavior. These assumptions facilitate large-scale data collection and reproducible experimentation by reducing the complexity of realistic browsing sessions. However, they often fail to capture the diversity, unpredictability, and temporal dynamics of real Tor users. For example, the closed-world assumption assumes that users visit only a predefined set of monitored websites known to the attacker. Under this setting, both the training and testing datasets consist exclusively of traffic traces collected from these monitored websites, ensuring that every testing instance belongs to one of the predefined classes. This considerably simplifies the classification task and often enables WF classifiers to achieve exceptionally high accuracy. Although the closed-world setting provides a standardized benchmark for comparing different WF techniques, it does not accurately represent real-world browsing environments. In practice, Tor users may access an effectively unlimited and continuously evolving collection of websites, the vast majority of which are unmonitored. Consequently, an attacker must not only identify monitored websites but also distinguish them from a much larger population of unknown webpages, making the classification task substantially more challenging~\cite{aminuddin2023rise}.

\subsubsection{Challenges in Genuine Dataset} Unlike synthetic datasets, genuine WF datasets are collected from real user browsing activities rather than controlled experiments. To date, GTT23~\cite{jansen2026measurement} is the only publicly available large-scale genuine Tor traffic dataset for WF research. The dataset is collected at a Tor exit relay, where encrypted traffic can be associated with destination domain names, enabling automatic traffic labeling without requiring active webpage crawling or manual annotation.

Compared with synthetic datasets, GTT23 significantly improves dataset realism by naturally incorporating many factors that are difficult to reproduce in controlled environments. Because the traffic is generated by real Tor users, the dataset inherently captures diverse browsing behaviors, heterogeneous system and network environments, browser caching effects, dynamic webpage content, advertisements, third-party resources, and other real-world variations. Consequently, it eliminates several simplifying assumptions commonly adopted during synthetic dataset construction, including the static webpage assumption, passive webpage assumption, cache-disabled assumption, and replicability assumption. As a result, GTT23 provides a more realistic benchmark for evaluating the robustness and practical applicability of WF attacks.

Despite these advantages, genuine datasets still present several unique challenges. First, GTT23 does not eliminate the \emph{traffic parsing assumption}, as individual page visits must still be accurately identified and segmented from continuous traffic traces. However, this assumption is unrealistic for a practical WF adversary observing traffic between the client and the Tor entry relay, where page boundaries are not explicitly available. The problem becomes even more challenging during concurrent browsing, as users frequently open multiple browser tabs simultaneously, producing naturally overlapping traffic traces whose boundaries are difficult to distinguish. Existing genuine datasets provide limited support for realistic multi-tab browsing, and the development of representative multi-tab datasets remains an important open research challenge.

Second, because GTT23 is collected at the Tor exit relay, the identity of the originating Tor client is unknown. Traffic traces associated with the same destination domain may therefore originate from many different users with diverse hardware platforms, browser configurations, network conditions, geographic locations, and browsing habits. While this diversity improves the realism of the dataset, it also introduces substantial intra-class variability that is absent in most synthetic datasets. In particular, timing-based features become significantly less stable because traces collected from the same website may be separated by arbitrary time intervals and generated under highly heterogeneous environments. Consequently, classifiers that heavily rely on packet timing characteristics may experience noticeable performance degradation when evaluated on genuine datasets.

Finally, the exit-relay collection methodology inherently limits the applicability of genuine datasets to user-specific WF research. Since individual Tor clients cannot be reliably identified or linked across browsing sessions, the collected traffic cannot accurately reflect a specific user's browsing history or behavioral patterns. As a result, the current genuine datasets are well-suited for studying population-level website fingerprinting. Still, they are less appropriate for evaluating attacks that require client-specific traffic modeling, user profiling, or personalized classifier adaptation.

\subsection{Future Research Directions}

Despite significant progress in WF research, a substantial gap remains between laboratory evaluations and real-world deployment. In this section, we organize future research directions into three categories: WF attacks, WF defenses, and WF datasets.

\subsubsection{WF Attacks}

\paragraph{Relaxing Unrealistic Assumptions.} Existing WF attacks commonly rely on assumptions that facilitate model development and evaluation but do not accurately reflect real-world browsing environments~\cite{deng2026enhancing}. Future research should systematically relax these assumptions by considering more realistic threat models, browsing behaviors, and network conditions, thereby enabling a more rigorous assessment of attack robustness and practical applicability.

From a user behavior perspective, future attacks should incorporate multi-tab browsing, asynchronous page loading, background application traffic, and diverse navigation patterns, all of which introduce substantial noise and traffic overlap. From an attacker capability perspective, future research should consider limited knowledge settings, where attackers have incomplete information about target websites and user environments, requiring uncertainty-aware and adaptive attack strategies. From a webpage perspective, evaluations should extend beyond static homepage traces to include dynamic content, personalized pages, cached pages, and interactive web applications.

\emph{Adaptive Website Fingerprinting under Distribution Shift.}
The performance of WF attacks deteriorates over time due to distribution shifts introduced by evolving website content, network dynamics, and user browsing behavior. Addressing this challenge requires adaptive learning paradigms that can continuously maintain model effectiveness under changing environments. Promising directions include online learning, continual learning, incremental learning, and domain adaptation to enable efficient adaptation to evolving traffic distributions. Furthermore, longitudinal datasets that capture the temporal evolution of websites are essential for systematically evaluating model robustness and adaptation capabilities. Incorporating ensemble learning and uncertainty-aware modeling may further enhance resilience to distribution shifts and improve the long-term reliability of WF systems.

\emph{Foundation Models for Website Fingerprinting.}
Recent advances in foundation models have opened new research opportunities for both WF attacks and defenses. Transformer-based architectures and large language models (LLMs) have demonstrated remarkable capabilities in modeling complex sequential and structured data, making them well suited for encrypted traffic analysis. By representing traffic traces as tokenized sequences or structured tokens, these models can learn discriminative traffic fingerprints from packet-level or burst-level observations while capturing long-range dependencies and global traffic characteristics that are challenging for conventional architectures to model~\cite{jiao2025llm,song2025redefining,zhao2025embracing,long2025survey}.

Future research may explore pretrained traffic foundation models that enable transfer learning across datasets, network environments, and attack scenarios. Other promising directions include prompt-based and in-context learning for efficient adaptation to emerging websites with limited training data, as well as multimodal foundation models that jointly exploit traffic statistics, temporal dynamics, and graph-based representations. Furthermore, integrating generative models and reinforcement learning with foundation models may facilitate the development of adaptive defense mechanisms that simulate adversarial behavior and optimize the trade-off between privacy protection and system utility.

\subsubsection{WF Defenses}

\paragraph{Practical and Efficient Defense Deployment.}
Although many WF defenses demonstrate strong effectiveness in experimental settings, their practical deployment remains limited by high latency, bandwidth overhead, and computational cost. Future research should focus on lightweight, scalable, and deployable defense mechanisms that provide robust privacy guarantees while preserving the low-latency and usability requirements of anonymous communication systems~\cite{khajavi2026lightening}.

\emph{Adaptive and Cross-Layer Defense Mechanisms.}
Future WF defenses may explore cross-layer designs that jointly leverage mechanisms at the network, transport, and application layers to improve both efficiency and robustness. Furthermore, adaptive defense strategies that dynamically adjust traffic perturbation according to network conditions, user behavior, and estimated threat levels may achieve more effective privacy–performance trade-offs than static defense mechanisms.

\emph{Robustness Against Adaptive and Reverse Engineering Attacks.}
Most existing WF defenses assume a fixed adversarial model and rarely consider attackers capable of reverse engineering the defense mechanism or exploiting knowledge of the underlying attack model~\cite{khajavi2026lightening}. Future research should investigate defense strategies that remain effective under adaptive adversaries with partial or complete knowledge of the defense algorithm, following Kerckhoffs's principle~\cite{petitcolas2011kerckhoffs}. Evaluating robustness against model-aware attacks, defense-aware training, and adaptive attack strategies will be essential for assessing the practical security of future WF defenses.

\subsubsection{WF Datasets}

\paragraph{Dataset diversity and realism.}
The quality and diversity of datasets fundamentally influence the development and evaluation of WF attacks and defenses. Future datasets should better reflect real-world browsing scenarios by incorporating internal webpages, dynamic web content, user interactions, multi-page navigation, and realistic browsing sessions, rather than relying primarily on homepage traces. In addition, open-world datasets should include large-scale unmonitored traffic and diverse background activities to enable more rigorous evaluation of false-positive rates, scalability, and practical deployment.

Although GTT23~\cite{jansen2026measurement} was collected from real user traffic at Tor exit relays, it does not eliminate the traffic parsing assumption. The captured traffic consists of continuous, interleaved flows from multiple users, making it challenging to accurately identify and isolate the browsing activities of individual users. Consequently, additional traffic segmentation and user attribution techniques are required before the dataset can be effectively used for website fingerprinting.

\emph{Temporal and longitudinal datasets.}
Addressing distribution shifts requires datasets that capture the temporal evolution of websites, user behavior, and network conditions. Future research should establish longitudinal datasets collected over extended periods to facilitate systematic evaluation of model robustness, adaptation capability, and long-term performance under evolving environments. Such datasets are essential for advancing adaptive and continual learning approaches for WF.

\section{Conclusion}
\label{sec:conclusion}

Despite the growing significance of WF attacks and defenses, a systematic and comprehensive survey of the field has remained lacking. This survey addresses this gap by presenting a unified taxonomy of Tor-based WF research, attack models, defense mechanisms, and encompassing datasets. We provide a comprehensive analysis of representative approaches, examining their design principles, threat assumptions, effectiveness, deployment trade-offs, and applicability across diverse evaluation settings. Beyond reviewing existing techniques, we systematically analyze the limitations and open challenges of current WF datasets, attacks, and defenses, including issues related to realistic browsing behavior, model drift, traffic segmentation, and deployment in practical environments. Finally, we identify promising research opportunities and emerging directions that can guide the development of more robust, adaptive, and deployable WF attacks and privacy-preserving defenses, providing a foundation for future advances in Tor security.

\bibliographystyle{ACM-Reference-Format}
\bibliography{references}


\end{document}